\documentclass[sigconf]{acmart}

\renewcommand\footnotetextcopyrightpermission[1]{} 

\def\BibTeX{{\rm B\kern-.05em{\sc i\kern-.025em b}\kern-.08emT\kern-.1667em\lower.7ex\hbox{E}\kern-.125emX}}

\usepackage{graphicx}

\usepackage{threeparttable}
\usepackage{multirow}

\usepackage{url}
\usepackage{hyperref}
\usepackage{breakurl}

\usepackage{tabularx}
\usepackage{bigstrut}
\usepackage{amsmath}
\usepackage{amsfonts}
\usepackage{amssymb}
\usepackage{wasysym}
\usepackage{enumitem}

\usepackage[british]{babel}

\renewcommand{\paragraph}[1]{\vspace{0.1in}\noindent{\bf{#1}.}}

\usepackage{ragged2e}

\begin{document}
\sloppy

\fancyhf{}

\title{Trick or Heat? Manipulating Critical Temperature-Based Control Systems Using Rectification Attacks}


\author{Yazhou Tu}
\authornote{Tu and Rampazzi are co-first authors.\\ Corresponding faculty authors: S. Rampazzi, K. Fu, X. Hei}
\affiliation{%
  \institution{University of Louisiana at Lafayette}
}
\email{yazhou.tu1@louisiana.edu}

\author{Sara Rampazzi}
\affiliation{%
  \institution{University of Michigan}
  }
\email{srampazz@umich.edu}
\authornotemark[1]

\author{Bin Hao}
\affiliation{%
  \institution{University of Louisiana at Lafayette}
}
\email{bin.hao@louisiana.edu}

\author{Angel Rodriguez}
\affiliation{%
  \institution{University of Michigan}
  }
\email{angelrod@umich.edu}

\author{Kevin Fu}
\affiliation{%
  \institution{University of Michigan}
  }
\email{kevinfu@umich.edu}

\author{Xiali Hei}
\affiliation{%
  \institution{University of Louisiana at Lafayette}
}
\email{xiali.hei@louisiana.edu}

\renewcommand{\shortauthors}{Tu and Rampazzi, et al.}

\begin{abstract}
Temperature sensing and control systems are widely used in the closed-loop control of critical processes such as maintaining the thermal stability of patients, {or in alarm systems for detecting \RaggedRight}
temperature-related hazards. However, the security of these systems has yet to be completely explored, leaving potential attack surfaces that can be exploited to take control over critical systems.

In this paper we investigate the reliability of temperature-based control systems from a security and safety perspective. We show how unexpected consequences and safety risks can be induced by physical-level attacks on analog temperature sensing components. For instance, we demonstrate that an adversary could remotely manipulate the temperature sensor measurements of an infant incubator to cause potential safety issues, without tampering with the victim system or triggering automatic temperature alarms. This attack exploits the unintended rectification effect that can be induced in operational and instrumentation amplifiers to control the sensor output, tricking the internal control loop of the victim system to heat up or cool down. Furthermore, we show how the exploit of this hardware-level vulnerability could affect different classes of analog sensors that share similar signal conditioning processes.

Our experimental results indicate that conventional defenses commonly deployed in these systems are not sufficient to mitigate the threat, so we propose a prototype design of a low-cost anomaly detector for critical applications to ensure the integrity of temperature sensor signals.
\end{abstract}


\maketitle
\section{Introduction}   

Embedded systems that utilize temperature sensors are extensively employed in the supervision and automatic control of temperature-sensitive environments such as in hospitals, laboratories, and industrial and manufacturing facilities \cite{thermocouple,rtdomega,feteira2009negative,antonucci2009infant}. In particular, closed-loop temperature control systems have become indispensable in many critical applications such as infant incubators that maintain the thermal stability of low birth weight or sick newborns \cite{who_infantdeath}, and blood bank or vaccine refrigerators that provide an optimal preservation temperature in the cold chain \cite{who_coldchain01,who_bloodcoldchain}.

In this paper, we present a research study on the reliability of temperature-based control systems and their sensors. Our study is driven by the importance of security in safety-critical temperature-based control systems and concerns about potential consequences caused by compromised sensors. It may not be safe to assume that these automatic systems would always behave as users expected or could always be carefully attended to by alert human operators. Moreover, some adverse effects caused by unsafe temperatures can be subtle and may not be detected immediately. We notice that there are reports about how safety issues can be related to improper temperature control \cite{overheat_death,molgat2013accidental,vaccine_spoil,who_bloodcoldchain02}. For instance, deaths and injuries to neonates in incubators have been linked to thermostat failure that caused incubator overheating and infant hyperthermia \cite{who_infantdeath}. In one case of a fatal incubator malfunction, an infant incubator overheated and resulted in the death of a baby \cite{overheat_death}. While the incubator's alarm went off, the nurses did not hear it because of the noisy, busy environment in the neonatal intensive care unit. Besides, poor refrigeration could make vaccines ineffective and leave the patients unprotected against dangerous diseases, or increase the risk of bacterial proliferation in blood products and cause potentially life-threatening transfusion reactions \cite{vaccine_spoil,vaccine_cdc,who_bloodcoldchain02}. Therefore, it is necessary to investigate and understand the security and reliability of critical temperature-based control systems.

\begin{figure}[t]
\centering
 \includegraphics[scale=0.74]{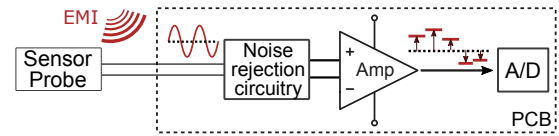}
  \vspace{-2mm}
  \caption{An illustration of the general signal conditioning path of a temperature sensor. Our attack can bypass conventional noise filtering and generate a controllable DC voltage offset at the ADC input. }\label{fig_conditioning_path}
 \vspace{-4mm}
\end{figure}

Our study focuses on physical-level attacks that exploit weaknesses in temperature sensors to manipulate temperature-based control systems. We show that, without tampering with the target system, adversaries can remotely manipulate the control system or circumvent temperature alarms by spoofing the temperature sensor with electromagnetic interference (EMI) signals. Unlike previous works that utilize the generation of subharmonics in non-linear circuit components to demodulate out-of-band EMI signals \cite{kune2013ghost}, or induce signal clippings in Electro-Static Discharge (ESD) protection circuits of a  microcontroller  \cite{selvaraj2018electromagnetic}, 
our attack exploits the unintended rectification effect in operational and instrumentation amplifiers to generate a controllable DC component on the amplifier output that can be used to manipulate the sensor readings (Fig. \ref{fig_conditioning_path}). We conduct detailed signal injection experiments on a typical temperature sensing circuitry and show that a stabilized voltage level can be intentionally induced and controlled to increase or decrease the sensor output. We analyze the vulnerability and attack surface of circuit components with both direct power injection (DPI) and remote signal injection experiments. We then investigate the effect of remote attacks on several off-the-shelf temperature sensors and control systems that use different amplifiers. In addition, we show how this physical-level exploit can affect other classes of sensors that share similar signal conditioning processes.

To explore potential consequences and understand the threats of physical-level attacks on critical temperature-based control systems, we study our attacks on an infant incubator and other real-world systems.
In particular, we show how an adversary can remotely manipulate an infant incubator temperature to cause life-threatening issues.
Without triggering the automatic temperature alarms, the attack can trick the internal control system of the infant incubator to heat the cabin up to $38.5 ^{\circ}C$ or cool it down to $29 ^{\circ}C$, from attack distances of 1.9 m and 1 m respectively in the open air with a transmitting power of 4 W. These dangerous temperatures can raise the risk of temperature-related health issues in infants, such as hyperthermia and hypothermia, which in turn can lead to hypoxia, neurological complications, and even death \cite{walter2016neurological,mance2008keeping,severe_hypo}. We also investigate the threats on several systems equipped with different types of temperature sensors such as laboratory thermal control equipment and 3D printers.
Our experimental results show that these systems blindly trust the spoofed temperature sensor readings, resulting in manipulated decision makings of the victim system.

Our study illustrates the threat of exploiting a low-level vulnerability of temperature sensors in critical control systems and the necessity to mitigate this vulnerability.
We discuss several conventional defenses, such as filtering and shielding, as well as their limitations. To enhance the robustness of critical temperature-based control systems and shed light on defenses against rectification attacks on sensors, we propose a low-cost anomaly detector that identifies malicious interference in the vulnerable frequency range. Once the interference is detected, the signal information can be used by the system software to estimate the sensor data reliability.
Our study aims to raise the awareness of potential threats of compromising temperature sensors and work towards improved security and resilience in future designs of critical temperature-based systems.

\renewcommand{\thefootnote}{\arabic{footnote}}

In summary, we list our main contributions as follows:
\vspace{-0.5mm}
\begin{itemize}[leftmargin=4mm]
\vspace{-0.5mm}
\item We investigate the reliability of temperature-based control systems and their sensors from a security and safety prospective. We explore how unexpected consequences can be caused in real-world systems with physical-level attacks on temperature sensors\footnotemark.

\renewcommand{\thefootnote}{\arabic{footnote}}

\footnotetext{Demo videos of the proof-of-concept attacks are available at \url{https://www.youtube.com/playlist?list=PLZaFM1g7JkPgpieNXMomMTQ7w9iZ8Yn-3}.}

\vspace{-0.5mm}
\item We bridge the gap of sensor security research by explaining how to manipulate the DC voltage of temperature sensor signals, characterizing the rectification effect that can be intentionally induced in amplifiers. By analyzing the attack surface of circuit components with DPI and remote EMI injection experiments, we unveil the fundamental causality of the vulnerability. Furthermore, we show that the exploit of the rectification phenomenon could affect other classes of sensors that use similar vulnerable components.

\vspace{-0.5mm}
\item  Based on the experimental results of our study, we discuss conventional defensive strategies, their limitations and challenges; then we propose a prototype design of an analog anomaly detector to enhance the security and reliability of temperature-based control systems.
\end{itemize}

\vspace{-2mm}
\section{Background}

In this work, we focus on the security of systems based on three types of analog temperature sensors: thermocouples, Resistance Temperature Devices (RTDs), and thermistors. Thermocouples operate on the Seebeck effect, which occurs when two dissimilar metals are joined at one end. The output voltage is a direct function of the temperature difference between the junction of the metals and the target measurement point \cite{thermocouple}. RTDs are constructed of a metal, such as copper or platinum, which increases in resistance with increasing temperature. Compared to thermocouples, they require voltage or current excitation, and are generally more sensitive. Finally, thermistors are made of metal oxides and may have either a negative or positive temperature coefficient. Negative temperature coefficient thermistors (NTCs) decrease in resistance with increasing temperatures, while positive temperature coefficient thermistors (PTCs) increase in resistance with increasing temperatures. Thermistors exhibit a much greater sensitivity than thermocouples or RTDs. However, their operating temperature range is narrower.

\vspace{-1mm}
\paragraph{Signal Conditioning of Analog Temperature Sensors}
Analog sensors require a signal conditioning phase before a data acquisition device can effectively process the signal.
Analog temperature sensors present specific signal conditioning requirements to provide reliable and accurate measurements.
For instance, the relationship between the output voltage and the temperature measurements is not linear, and each type of sensor exhibits its distinctive non-linearity. For this reason, analog temperature sensors often require high-resolution ADCs to achieve the desired accuracy \cite{kester1999section}.
Also, thermocouples require an additional correction to the acquired measurement called Cold-junction compensation (CJC). CJC accounts for the voltage offset generated at the connection between the thermocouple and the terminals of the acquisition device. In comparison, RTDs are often placed in bridge circuits for detecting small resistance changes. These additional considerations are used to improve the measurement accuracy.

Furthermore, because of the low-level voltage, the output signal from analog temperature sensors needs to be properly filtered and amplified before further processing can occur (Fig. \ref{fig_conditioning_path}).
RTDs and thermistors voltage outputs are usually amplified by operational amplifiers (op-amps), while thermocouples use instrumentation amplifiers (in-amps) \cite{kitchin2004designer}.
Both types of amplifiers provide the very important function
of extracting the small signals from the temperature sensors, and also providing the adequate common-mode noise rejection\footnotemark. Filters, on the other hand, block out both common and differential-mode noise, and the interference induced by the 50/60 Hz power.

Inadequate design specifications of these fundamental components can be exploited by an adversary to gain control over the sensor and induce the target system to make automated decisions based on untrustworthy sensor data.

\footnotetext{Depending on the conduction mode, differential-mode (or normal-mode) noise appear across the lines of an electric circuit following the same direction as the power supply current. In contrast, in common-mode noise, current flows in the same direction along different lines with the same voltage with reference to the earth \cite{common}.}

\vspace{-1mm}
\paragraph{Rectification Effect in Amplifiers}
The rectification effect in amplifiers is a phenomenon that converts AC signals in input to an amplifier to a DC offset component in the output signal. This offset is the result of the non-linear voltage-current characteristics of the internal diodes formed by silicon p-n junctions inside the transistors (FETs or BJTs) that constitute the amplifier internal input stage ~\cite{rfi_rectify_mt96, fiori2015analog,fiori2016senosr_sign, wu2018characterization}.
Generally, the operating point of an amplifier, also known as quiescent point, is the DC bias required by an amplifier to operate correctly and amplify the input signal without distortion. Especially in low-power amplifiers, where the input stage transistors works at low current and low impedance levels, a high frequency sine wave injection can alter the bias level of the amplifier, generating a DC offset in the output signal.

For example, considering a small AC voltage $V_{x}$ at frequency $\omega_{x}$ injected across the base-emitter junction $\Delta V = V_{x}cos(\omega_{x}t)$ of an operational amplifier BJT-based input stage,  the collector current around the quiescent point can be express as ${I_{C}'=I_{C}(V_{BE}+\Delta V)}$ where $V_{BE}$ is the base-emitter voltage. Applying the Taylor series expansion of the transistor collector current we can observe three main spectral components: the quiescent  collector current $I_{C}$, $cos(\omega_{x}t)$ and $cos^2(\omega_{x}t)$. While the linear spectral term is filtered by other stages within the device, the quadratic term remains and contains two components, one depended by twice of the signal input frequency ($2\omega_{x}$) and a DC term~\cite{rfi_rectify_mt96}. This DC term is the rectification effect, that can be expressed as a variation of the quiescent collector current:

\begin{equation}\label{eq_dc_offset}
\vspace{-1mm}
\begin{small}
\begin{array}{l}
\Delta i_{C} = (\frac{V_{x}}{V_T})^2 \cdot \frac{I_{C}}{4}
\end{array}
\end{small}
\vspace{-1mm}
\end{equation}

 where $V_T$ is a constant equal to 25.68 mV at 25 $^\circ C$ for BJT based amplifiers \cite{rfi_rectify_mt96}.
 In FET-input operational amplifiers the rectification term of the drain current $I_{D}$ become $\Delta i_{D} = (\frac{V_{x}}{V_P})^2 \cdot \frac{I_{DSS}}{2}$ where $I_{DSS}$ is the drain current at zero gate-source voltage, and $V_P$ the pinch-off voltage.

 The analysis shows how the rectification effect in op-amps is directly proportional to the square of the injected AC signal's amplitude, independently by the type of transistor used \cite{rfi_rectify_mt96}.

In addition, instrumentation amplifiers are generally composed by three op-amps, where the first two are arranged to buffer each input to the third one. Wu et al. \cite{wu2018characterization} demonstrated that the rectification effect mainly happens at the non-inverting input of two op-amps in the first stage of an in-amp. Furthermore, the resulting DC offset at the in-amp output increases if the DC voltage difference between the inverting input and the non-inverting input of the third op-amp becomes higher.
Therefore, to reduce the rectification effect, external noise signals should be eliminated before the amplifier input with proper filtering.

Our remote attack targets temperature-based control systems lacking effective noise suppression circuits, tuning the transmitted EMI signals to a carrier frequency equal to the resonant frequency of the target circuit component to maximize the injected AC voltage and induce the rectification effect.

\vspace{-1mm}
\section{Threat Model}
\vspace{-0.5mm}

The goal of the adversary is to spoof the temperature sensor measurement and manipulate a temperature control system to heat up or cool down to an unsafe temperature. The adversary cannot tamper with any hardware or software of the target system. Also, we don't consider a malicious human operator that could directly affect the actual temperature around the sensor or deliberately operate the victim system to manipulate the temperature setpoints of the system.

\vspace{-4.5pt}
\paragraph{Attack Scenarios}
Adversaries could launch the attack from one to several meters away, depending on their equipment and susceptibility of the victim system. Furthermore, the malicious EMI signals can penetrate many common physical barriers such as walls and windows. For instance, the attack could be launched from outside of a wall/window or from adjacent rooms.
An adversary could also use a hand-held attack device that can be carried and surreptitiously operated under his/her clothes. Additionally, the adversary might secretly leave or install a small remote control EMI emitting device around the victim system in advance of the attack. During the attack, two parameters (frequency and amplitude of EMI signal) need to be adjusted.

\vspace{-4.5pt}
\paragraph{Equipment}
Adversaries could use commodity signal generators, amplifiers and antennas to emit malicious EMI signals. Alternatively, adversaries could purchase or make a customized small portable transmitter to conduct the attack; the device would be similar to a hand-held radio transmitter (e.g., walkie-talkie) but with gain control and a frequency range that covers the attack frequencies. The power of EMI emitters that we use in experiments is below 4 W, but more capable adversaries might use more specialized equipment and techniques to improve the attack.

\vspace{-4.5pt}

\paragraph{Feedback} We assume that the adversary can observe the temperature readings or heating/cooling indicator lights in the target system, to ensure the induced attack effect. Alternatively, another adversary or a monitoring device could help observe the feedback of the victim system. However, the adversary does not have to observe the victim system all the time; after the adversary ensures the attack effect and selects suitable frequencies and amplitudes of attack signals, observations will no longer be needed.

\vspace{-2mm}
\section{Compromising Temperature Sensors}\label{sec_sensors}
\vspace{-1mm}

In this section, we conduct detailed signal injection experiments on typical temperature sensing circuits to study the attack effect. We analyze the vulnerable circuit components and attack surface with both DPI and remote EMI injection experiments.


\vspace{-2mm}
\subsection{Security Analysis}
\vspace{-1mm}

\begin{figure}[t]
\centering
 \includegraphics[scale=0.59]{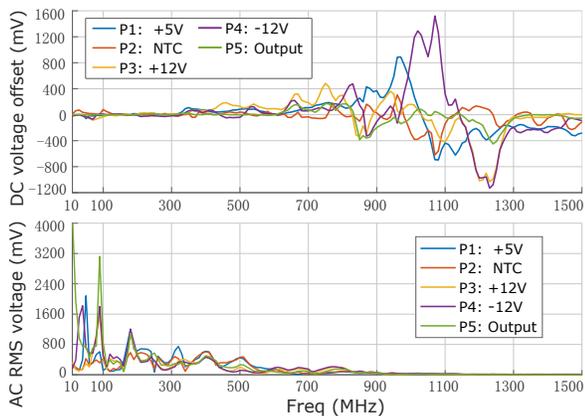}
  \vspace{-4mm}
  \caption{The results of DPI experiments on different injection points of the experimental circuitry. We record the induced DC voltage offset and the RMS voltage of AC signals corresponding to different EMI frequencies.
  }\label{fig_dpi_lm1458_freq}
 \vspace{-5mm}
\end{figure}

To explain how temperature sensors could be affected by rectification attacks, we build an experimental temperature sensing circuitry based on an NTC thermistor.
We wire the thermistor in a bridge circuit.
Bridge circuits are commonly used in the wiring of resistive sensors such as thermistors, RTDs and strain gauges \cite{kester1999practical}. The differential voltage generated by the bridge circuit is collected and amplified by a Texas Instruments (TI) LM1458 operational amplifier. The details of the setup can be found in the Appendix (Fig. \ref{fig_ntc_circuits}).

\vspace{-4pt}
\paragraph{Direct Power Injection (DPI) Experiments} It is difficult to measure and analyze the exact attack effect in circuits caused by remote EMI radiations since the path and strength of the induced EMI signals cannot be accurately predicted. Thus, we conduct DPI experiments to identify and analyze how EMI can affect internal components of temperature sensors.

In DPI, EMI signals can be injected directly into desired injection points on the circuit through conductance. In this way, we can control the power of the injected EMI signals more accurately and avoid interference from unintentional EMI radiations on other parts of the circuits. In our experiments, we connect the direct power injection circuit to each of the possible signal injection points on the sensing circuitry.

\paragraph{Inducing a Stabilized DC Voltage with Specific EMI Signals}
To achieve adversarial control over the sensor output instead of general disruptions of the sensing system, we need to induce stabilized DC voltage levels to control the sensor output rather than fluctuating interference signals to disturb it. First, we find specific EMI signals that can be rectified by the amplification circuits to induce controllable voltage levels without causing strong noises. We inject single-tone sine-wave EMI signals to each injection point of the experimental circuitry and sweep the frequency from 10 MHz to 1.5 GHz at an interval of 10 MHz with an injection power of 15 dBm, which is equivalent to 32 mW. As shown in Fig. \ref{fig_dpi_lm1458_freq}, we record the induced DC voltage offset as well as the root mean square (RMS) voltage of fluctuating alternating current (AC) signals in the output of the amplifier. We observe that EMI signals at specific frequencies induce a significant DC offset and the corresponding AC interference signal is below the typical noise floor. Such frequencies can be used in attacks to induce intentionally fabricated signals that cannot be easily distinguished from legitimate sensor measurements. Depending on the frequency of EMI signals, the induced DC offset in the experimental circuitry could be either positive or negative, allowing adversaries to increase or decrease the temperature measurement maliciously.

\begin{figure}[t]
\centering
 \includegraphics[scale=0.58]{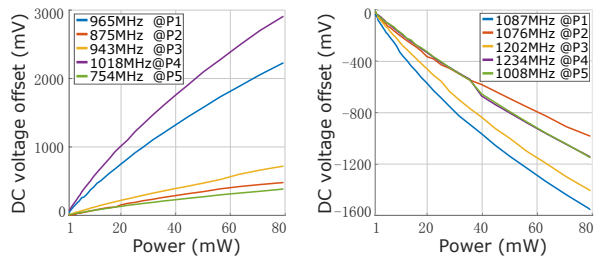}
  \vspace{-3mm}
  \caption{The relationship between the magnitude of the induced DC voltage offset and the power of directly injected EMI signals. }\label{fig_dpi_lm1458_mw}
 \vspace{-4mm}
\end{figure}

\vspace{-2pt}
\paragraph{Attack Surface} The identification of the attack surface helps to understand possible attack mechanisms and facilitates the evaluation of sensor security in future system designs.
As shown in Fig. \ref{fig_dpi_lm1458_freq}, our DPI experiments validate that a stabilized DC voltage signal can be induced by EMI signals injected through different entry points, including the sensor wire as well as other parts of the circuitry such as shared power lines. As a result, adversaries could exploit sensor wires, relatively long cables or printed circuit board (PCB) traces to inject malicious EMI signals and induce the attack effect. The potential attack surface also includes other components in the system that are connected to the injection points of the sensing circuitry. For instance, EMI signals conducted through the charging port could affect a physically co-localized microphone in a smartphone \cite{kasmi2018remote}. Similarly, devices, cables and other components that are connected to possible injection points of the temperature sensing circuitry could also make the sensor more susceptible to attacks and need to be considered in the design of a system.

\paragraph{DC Voltage and EMI Power Relationship} Adversaries need to control the magnitude of the induced DC voltage to gain effective control over the sensor output. Theoretically, the magnitude of the induced DC voltage offset is directly proportional to the power of injected EMI signals as described in Eq. (\ref{eq_dc_offset}).
Therefore, in the case of bipolar junction transistor (BJT) based amplifiers, the rectified DC current change $\Delta I$ can be described as
$\Delta I = (\frac{V_{emi}}{V_T})^2 \cdot \frac{I_C}{4}$, where $V_{emi}$ is the amplitude of injected EMI signals,
$I_C$ is the quiescent collector current of the transistor, and
$V_T$ is a constant.
Assuming that the equivalent resistance of the receiving circuitry is $R_r$, the power of the injected EMI is $P_r$, we have $V_{emi} = \sqrt{P_r R_r}$. Therefore, we can represent the induced DC offset as

\vspace{-2mm}
\begin{equation}\label{eq_dc_power}
\begin{small}
\begin{array}{l}
\Delta V_{DC} = \Delta I R_r = (\frac{V_{emi}}{V_T})^2 \cdot \frac{I_C}{4}R_r =  (\frac{R_r}{V_T})^2 \cdot \frac{I_C}{4}P_r
\end{array}
\end{small}
\end{equation}

\begin{figure}[t]
\centering
 \includegraphics[scale=0.63]{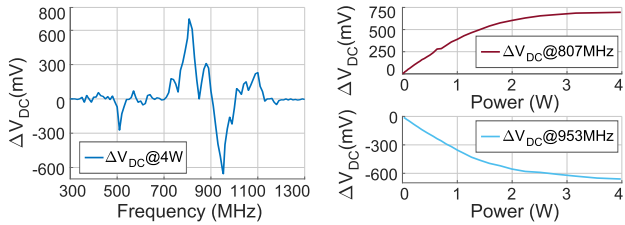}
   \vspace{-3mm}
  \caption{The relationship between the induced DC voltage offset and the attack frequency (left), and the relationship between the magnitude of the DC voltage offset and the transmitting power (right) in remote attacks.}\label{fig_ntc_radiation}
  \vspace{-3mm}
\end{figure}

We conduct DPI experiments on the experimental circuitry and inject EMI signals to each of the injection points to validate the effectiveness of the theoretical analysis. We select the EMI frequencies that correspond to peaks and troughs in Fig. \ref{fig_dpi_lm1458_freq} to affect the output of the amplifier. As shown in Fig. \ref{fig_dpi_lm1458_mw}, the power of directly injected EMI signals is positively related to the magnitude of the induced DC offset. The relationship can be considered as locally proportional but presents a changing rate that gradually decreases as the power of injected EMI signals grows.

\begin{figure}[t]
\centering
 \includegraphics[scale=0.65]{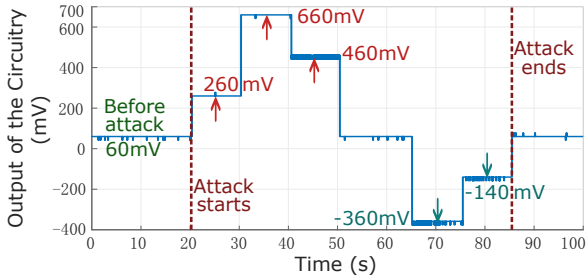}
   \vspace{-3.5mm}
  \caption{Remotely injecting stabilized voltage levels to control the output of the temperature sensing circuitry.}\label{fig_ntc_control}
  \vspace{-4mm}
\end{figure}

For simplicity, we utilize the free space propagation model to
understand the relationship between the transmitting power ($P_t$) and the injected power ($P_r$) in remote attacks. From the Friis transmission equation, we have

\vspace{-2.5mm}
\begin{equation}\label{eq_power_trans}
\begin{array}{l}
P_r = G_t G_r (\frac{\lambda}{4 \pi D})^2 P_t
\end{array}
\end{equation}

$G_t$ and $G_r$ are the gains of the transmitting and receiving antennas respectively. $G_t$ depends on the type of antenna that is used by the attacker. Note that components in the victim circuit work as a receiving antenna. $\lambda$ is the wavelength of EMI signals. $D$ is the attack distance between the adversary's antenna and the victim circuit. Based on Equations \ref{eq_dc_power}, \ref{eq_power_trans}, and our previous analysis, we can infer that the magnitude of the induced DC voltage signal is locally proportional to the power of transmitting EMI, which will be validated in our remote EMI injection experiments.

\vspace{-1mm}
\paragraph{Spoofing the Temperature Sensor Output}
We investigate remote EMI injections that leverage the rectification effect in amplifiers to gain adversarial control over the output of the temperature sensing circuitry.
As shown in Fig. \ref{fig_ntc_radiation}, we transmit single-tone sine-wave EMI signals and sweep the frequency from 300MHz to 1 GHz at an interval of 10 MHz with a transmitting power of 36 dBm (equivalent to 4 W) and observe the induced DC voltage offset on the oscilloscope. We find that the maximum and minimum DC voltage offsets are induced at around 810 and 950 MHz respectively. We then adjust the frequency of EMI signals with an interval of 1 MHz to find the most effective frequencies that can be used in remote attacks to maliciously increase or decrease the output voltage of the circuitry. During the experiments, we shield the PCB with a metal box and cover the probe of the oscilloscope with aluminum shielding sleeves to mitigate unintentional interference. We aim EMI signals to the sensor wire with a directional antenna \cite{setup_ant} from a horizontal distance of 0.2 m.

We demonstrate how adversaries can intentionally induce stabilized voltage levels to control the output of the temperature sensing circuitry by remote rectification attacks (Fig. \ref{fig_ntc_control}). In the experiment, we increase the output of the circuitry by using an attack frequency of 807 MHz and decrease it with a frequency of 953 MHz. We manipulate the magnitude of the injected DC voltage level by adjusting the transmitting power between 0 and 2 W at an attack distance of 0.2 m. We monitor the real-time analog output of the circuitry with the oscilloscope and record it with an Arduino UNO R3 microcontroller that is connected to a laptop.

\vspace{-2mm}
\subsection{Off-The-Shelf Temperature Sensors}
\vspace{-1mm}

\begin{figure}[t]
\centering
 \includegraphics[scale=0.60]{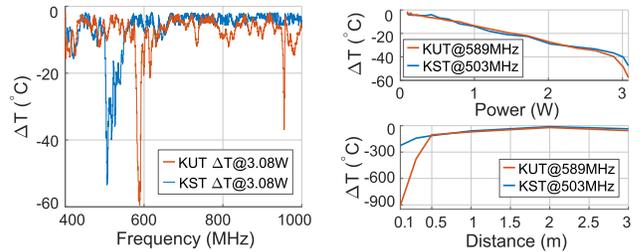}
   \vspace{-3mm}
  \caption{Results of remote attack experiments on K-type shielded (KST) and unshielded (KUT) thermocouples connected to the MAX31855K amplifier with an attack distance of 3 m in the open air (left and right top). The induced temperature change in different attack distances with a transmitting power of 3.08 W (right bottom).
  }\label{fig_ldc4ku}
   \vspace{-3mm}
\end{figure}

We investigate the attack effect on several off-the-shelf temperature sensor circuits that use different amplifier breakout boards.

\paragraph{Thermocouple Sensors}
We investigate the attack effects on both shielded and unshielded K-type thermocouples that are connected to a Sparkfun MAX31855K amplifier breakout board \cite{spark_31855} with a digital output interface, and an Adafruit AD8495 amplifier breakout board \cite{ada_8495} that has an analog output interface.

\begin{figure}[t]
\centering
 \includegraphics[scale=0.69]{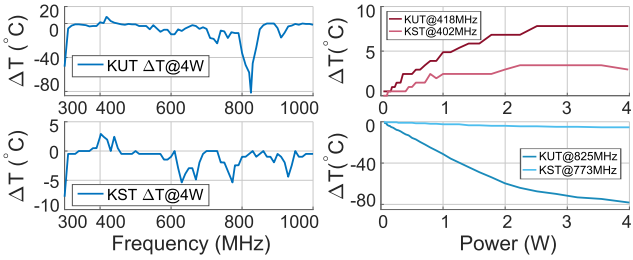}
   \vspace{-3mm}
  \caption{Results of remote attack experiments on thermocouples connected to the AD8495 amplifier with an attack distance of 0.6 m.}\label{fig_TC_AD8495}
   \vspace{-4mm}
\end{figure}

The length of the thermocouples we test is 1 m and we use an Arduino board to sample the output of the Sparkfun MAX31855K breakout board. As shown in Fig. \ref{fig_ldc4ku}, with an attack frequency of 589 MHz and an emitting power of 3.08 W, the attack can decrease the temperature measurement of the unshielded thermocouple by about $56 ^{\circ}C$ or $909 ^{\circ}C$ from an attack distance of 3 m or 0.1 m respectively. We also conduct the remote attack experiments on the Adafruit AD8495 breakout board using a similar setting and summarize the results in Fig. \ref{fig_TC_AD8495}.

\begin{figure}[t]
\centering
 \includegraphics[scale=0.58]{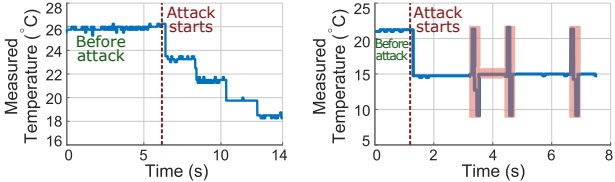}
   \vspace{-4mm}
  \caption{Remote control of a K-type shielded thermocouple at 1 m distance in two different scenarios: generating a step function (left) and spelling of the word ``HI" (right).}\label{fig_walnut}
   \vspace{-4mm}
\end{figure}

\vspace{-1mm}
\paragraph{Spoofing Attacks on Thermocouples}
Adversaries that have capabilities to deliver EMI to a victim thermocouple sensor circuitry can remotely spoof the sensor output and inject arbitrary, attacker-chosen temperature values.
We remotely inject spoofed temperature measurements to a K-type shielded thermocouple that is connected to the MAX31855K amplifier with an attack distance of 1 m and a transmitting power below 3.08 W. Our experiments demonstrate the control over the temperature sensor output in two different scenarios (Fig.~\ref{fig_walnut}). We use amplitude-modulated EMI signals to control the sensor measurements. We assume a sinusoidal carrier signal $c(t) = A(t) \cdot \sin(2\pi f t)$, where $A$ is the amplitude of the signal, $t$ is the time, and $f$ represents a frequency that induces a DC voltage offset in the output of the victim circuitry. We vary the amplitude $A$ over time, according to the different scenarios.

\begin{table*}
\centering
\begin{threeparttable}
\caption{Results of attack experiments on real-world temperature-based control systems
} \label{table_closed_loop} \vspace{-5mm}
\begin{tabular}{c|c|c|c|c|c}
  \hline
  \multirow{2}{*}{\bf{Device}} & \bf{Sensor}$^\dagger$ & \multirow{2}{*}{\bf{Applications}} & \bf{$\Delta T_{Max@0.1m} (^\circ C)$} & \bf{$\Delta T_{Min@0.1m} (^\circ C)$} &  \bf{Max. Attack} \\
  & \bf{Type} &  & \bf{/Freq. (MHz)} & \bf{/Freq. (MHz)} & \bf{Distance$^\ddag$(m)}  \\
  \hline \hline
Air-Shields Isolette C100 Incubator
 & NTC & Medical Device & +58.4/530 & -15.9/847 &  5.8 $^*$  \\
\hline
  Fisherbrand Traceable Thermometer
 & NTC & Biomedical, Lab   & +37/690 & -22/730 &  3.4 $^*$ \\
\hline
 Thomas Traceable Thermometer
 & NTC & Biomedical, Lab   & +16/640 & -50/830 &  1.6 \\
\hline
UVP HB-500 Hybridization Oven
 & PTC &   Laboratory   & +42.4/516 & -2.8/453 &  3.3 \\
  \hline
Revolutionary Science Incufridge  & Un   & Laboratory  & +0.9/308 & -3.3/309 &  0.6\\
  \hline
Sun Electronic EC12 Thermal Chamber & KTC  & Manufacturing, Lab   & +3.35/686    & -2.88/1300  & 0.3   \\
  \hline
MakerBot 3D printer Smart Extruder +  & KST  & Manufacturing, Lab   & +10/1000    & N/A  & 0.25   \\
\hline
Inkbird ITC-100VH Controller & KST & IoT  &     $>$+78/556 & N/A  &  11.5 $^*$\\
  \hline
Inkbird ITC-1000F Controller & NTC &  IoT &     N/A  &  -10.6/713 &   0.9  \\
  \hline
Inkbird ITC-100RH Controller & RTD &  IoT  &  $>$+80.9/453     &  N/A &  16.2 $^*$  \\
\hline
\end{tabular}
    \begin{tablenotes}
      \small
      \item  $\dagger$ NTC/PTC: NTC/PTC thermistor, KTC: K-type thermocouple, KST:  K-type shielded thermocouple, Un: Unknown.
      \item $\ddag$ The maximum distance that we could induce a change of $0.5^\circ C$ in the temperature measurement with a transmitting power of 4 W. \quad $*$ Estimated.
      \vspace{-4mm}
    \end{tablenotes}
\end{threeparttable}
\end{table*}

\begin{figure}[t]
\centering
 \includegraphics[scale=0.59]{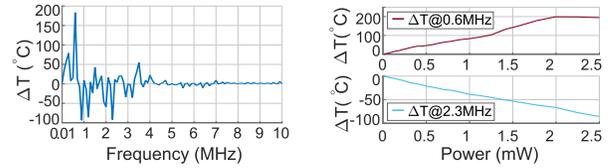}
  \vspace{-3mm}
  \caption{DPI experiments on the RTD circuitry with an injection power of 2.5 mW (left). The amount of induced temperature change with different injection power (right). }\label{fig_RTD}
   \vspace{-5mm}
\end{figure}

\vspace{-2mm}
\paragraph{Experiments with RTDs}
We test both shielded and unshielded PT100 RTDs connected to an Adafruit MAX31865 amplifier breakout board \cite{ada_31865} with remote EMI injection experiments. First, we generate EMI signals with antennas and sweep the frequency from 10 MHz to 1.5 GHz but could not observe a stable temperature change induced in the output of sensor.
We then inject conducted EMI signals directly into the terminals of the MAX31865 board connected to the RTD and sweep through a wider frequency range. As shown in Fig. \ref{fig_RTD}, we find that EMI signals with lower frequencies around 1 or 2 MHz can be more effective to manipulate the temperature measurement.

\vspace{-2mm}
\section{Manipulating Temperature-based Control Systems}
\vspace{-1mm}

In this section, we investigate the threats of the attack on real-world temperature-based control systems that use different kinds of temperature sensors, including NTC/PTC thermistors, thermocouples and RTDs. We evaluate the attacks on systems that are employed in medical applications such as an infant incubator, and in laboratory equipment that control critical biological experiments or manufacturing processes. Additionally, we investigate several commodity PID controllers equipped with temperature alarm functions.

We summarize the results of our attack experiments in Table~ \ref{table_closed_loop}.
We show that embedded systems based on different kinds of temperature sensors employed in different application areas can be affected by our attacks.
Our results validate that temperature-based control systems blindly trust temperature sensor readings to make automated decisions, which allows adversaries to exploit and abuse them for causing unintended consequences.

Many of the systems we test have external temperature sensor probes that need to be deployed to measure the temperature of a specific environment. Usually, the wiring and interfaces of systems with external sensors could make the system more susceptible to our attacks. Devices with internal temperature sensors might be less susceptible but can still be affected. For instance, the UVP HB-500 hybridization oven is covered by metal panels and most part of the internal sensor wire is protected by additional aluminum foil, but we notice that small gaps between the metal panels could allow EMI signals to pass through and be picked up by internal cables or PCB traces. In addition, control panels of many devices can allow EMI signals to enter the system. The control panels consist of various user interface components such as screens, buttons and lights; EMI signals could pass through the spaces between these components. Moreover, the cables connected to components in the control panel or peripheral devices could also pick up EMI signals and might conduct the signals into possible injection points of the victim temperature sensing circuitry.

\paragraph{Experimental Settings} The maximum transmitting power of our equipment is 36 dBm, which is  equivalent to 4 W.
We use a ZHL-4240 amplifier that has an average gain of about 44 dB in the range of 10 MHz to 2 GHz \cite{setup_amp}.
The signal source is an Agilent N5172B vector signal generator.
We use a directional antenna \cite{setup_ant} that has a length of 0.5 m to emit sinusoidal EMI signals with frequencies above 300 MHz, and an extendable dipole antenna for frequencies below 300 MHz.
For most of the devices, we sweep through 300 MHz to 1 GHz with an interval of 10 MHz and observe the temperature measurement of the device to find the attack frequencies. We then adjust the frequency with a step of 1 MHz to find the optimal attack frequency. If we could not find the attack frequencies for a device in this range, we would sweep through the frequency ranges of 10 to 300 MHz and 1 to 2 GHz. In Table \ref{table_closed_loop}, we record the maximum increase or decrease that can be induced in the temperature measurement of the target system and corresponding EMI frequencies with an attack distance of 0.1 m. For the Inkbird ITC-100VH and ITC-100RH controllers, the manipulated temperature can exceed the maximum temperature range of the device at an attack distance of 0.1 m.
We also record the maximum horizontal distance between the antenna and the target device that a change of $0.5 \ ^\circ C$ can be induced in the temperature measurement.
For several devices, the maximum attack distance is out of the dimension of our room setup, so we estimate the maximum distance based on our indoor measurements and the relationship between the induced temperature change and the attack distance (From Equations \ref{eq_dc_power} and \ref{eq_power_trans}, we have $\Delta V_{DC} \propto \frac{1}{D^2}$).

\vspace{-6pt}
\subsection{Medical Applications}
\vspace{-2pt}
\subsubsection{Infant Incubator}

Newborn infants regulate body temperature much less efficiently than adults \cite{who_thermoregulation}.
Infant incubators are critical medical devices widely used in neonatal care units. These incubators help maintain the thermal stability of infants - especially preterm or sick newborns \cite{who_infantdeath,survival_rate1,antonucci2009infant}. The temperature inside the cabin of incubators is measured and adjusted, via a closed-loop temperature control system, to reside within an ideal preset temperature range, minimizing the risks of morbidity and mortality \cite{walter2016neurological,mance2008keeping,severe_hypo}.

To maintain the infant in a Neutral Thermal Environment (NTE \cite{nurse2}), the closed-loop temperature control system in incubators can operate in skin servocontrol mode (skin-mode) or air temperature control mode (air-mode). The skin-mode is designed to maintain the neonate's abdominal skin temperature constant, whereas the air-mode is based on the control of the circulating incubator air temperature \cite{decima2013does}. The simplest way to achieve a thermoneutral environment is to maintain a constant abdominal skin temperature between $36.0 ^{\circ}C$ and $36.5 ^{\circ}C$, in the skin-mode. This range minimizes the number of calories needed to maintain normal body temperature and reduces the risks of cold stress or overheating  \cite{bell2006iowa_servocontrol}. Usually, NTC thermistors are used in infant incubators to measure the skin or air temperature and provide real-time feedback to the closed-loop temperature control system.

To find out whether the temperature control system of an infant incubator can be maliciously controlled and abused by adversaries to cause safety issues, we investigate our attacks on an Air-Shields Isolette C100 infant incubator \cite{incubator1}. We observe that the chassis of the incubator is shielded with aluminum panels. However, the large control panel, sensor interfaces, and air circulation holes on the chassis could still allow EMI signals to enter and affect the internal system components.
In our experiments, we aim the antenna to the front control panel of the infant incubator. However, attacks from other directions are also possible (e.g., targeting the back of the chassis from an adjacent room).

\begin{figure}[t]
\centering
 \includegraphics[scale=0.63]{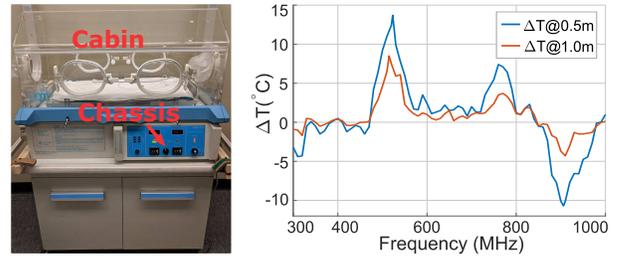}
 \vspace{-3mm}
  \caption{Infant incubator (left). The relationship between the induced change in the skin temperature measurement of the incubator and the attack frequency with a transmitting power of 4 W (right).}\label{fig_incubator_freq} \vspace{-4mm}
\end{figure}

\begin{figure}[t]
\centering
 \includegraphics[scale=0.68]{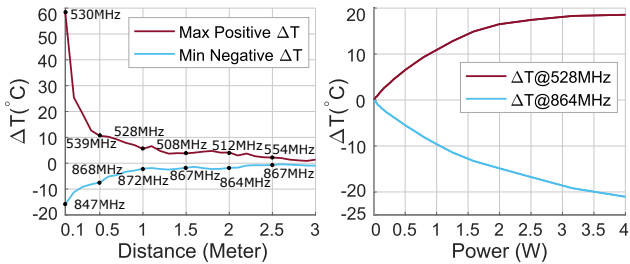}
 \vspace{-3mm}
  \caption{Maximum increase/decrease in the skin temperature measurement that can be achieved with different attack distances using a transmitting power of 4 W (left). The relationship between the amount of induced changes in the measured skin temperature and transmitting power with an attack distance of 0.2 m (right).}\label{fig_incubator} \vspace{-4mm}
\end{figure}

Using a transmitting power of 4 W, our attack can maliciously control the skin temperature sensor measurement of the infant incubator with certain attack distances. As shown in Fig. \ref{fig_incubator_freq}, an adversary can increase the skin temperature measurement by $8.5^\circ C$ or decrease it by $4.3^\circ C$ from 1 m away with attack frequencies of 515 MHz and 910 MHz respectively. Additionally, the air temperature sensor measurement of the incubator could also be affected by the attack, but the amount of the induced change is less significant (about $1.5^\circ C$ at an attack distance of 0.2 m). To understand possible attack distances that can cause safety threats in the incubator with a certain transmitting power, we measure the maximum increase and decrease that can be achieved with different attack distances using a transmitting power of 4 W (Fig. \ref{fig_incubator}). We observe that when we change the distance, the optimal attack frequency deviates slightly within a range of several tens of MHz. This could be caused by environmental changes when we change the distance. For instance, transmitting paths of reflected signals in the experimental area might have changed; and conductivity of objects nearby might affect the radiation pattern and impedance of the antenna. We also measure the relationship between the amount of induced changes in the measured skin temperature and transmitting power (Fig. \ref{fig_incubator}). The relationship is consistent with the results of our experiments on temperature sensor circuitry in Section \ref{sec_sensors}.

\vspace{-6pt}
\paragraph{Temperature Alarms} During the experiments, the incubator is functioning in the skin-mode. We notice that when the manipulated skin temperature measurement significantly deviates from the preset skin temperature, an alarm would be triggered. The incubator system continuously compares the skin temperature measurement with the preset temperature value and raise the preset temperature alarm when a difference larger than  $1^\circ C$ is detected \cite{incubator1}.

Additionally, a high temperature alarm would be triggered if the air temperature is over $38.5^\circ C$. The alarm system of the incubator continuously monitors the measurement of an extra internal high air temperature sensor and raises the high temperature alarm when the temperature exceeds the maximum temperature limit. The high temperature limit is $38.5^\circ C$ in the Air-Shields C100 incubator \cite{incubator1}, and could be higher in other systems \cite{incubator3}.
Finally, there is also a probe alarm function that detects shorted, open or disconnected conditions in air, skin, or high temperature sensors. However, the temperature alarm systems in incubators may not perform safety precautions reliably if the security of the system is compromised with physical-level attacks on temperature sensors. As a result, adversaries can manipulate the infant incubator system to cause safety issues without triggering any of these alarms.

\vspace{-1.5mm}
\paragraph{Heating Attacks} An adversary can decrease the skin temperature measurement maliciously and trick the internal control loop of the incubator to actuate the heating system. With an attack distance of 2 m, an adversary that uses a transmitting power of 4 W can decrease the measured skin temperature by $1.8^\circ C$ (Fig. \ref{fig_incubator}). Adversaries can also launch the attack from an adjacent room. In our experiments, the infant incubator is placed 0.1 m away from a wall that has a thickness of 0.15 m and we target the back of the chassis from an adjacent room.
With the wall between the adversary and the incubator, attacks with the same transmitting power can decrease the skin temperature measurement by $4.5^\circ C$. As a result, the system would try to compensate for the induced temperature change to maintain the ``preset temperature" by actuating the heaters.

To avoid being detected by the preset temperature alarm, adversaries can increase the transmitting power slowly and maintain the difference between the measured and preset temperature less than $1^\circ C$. Adversaries can manipulate the system to reach and keep the maximum temperature of $38.5^\circ C$ without triggering the high temperature alarm. This excessive temperature can result in hyperthermia in newborns with consequent dehydration, lethargy, seizures, apnea, increased risks of neurologic injury, etc. \cite{severe_hypo,kasdorf2013hyperthermia}.

\vspace{-1.5mm}
\paragraph{Cooling Attacks} There is no automatic alarm to be triggered in the incubator if the cabin temperature drops below a specific minimum threshold. As a result, with an attack distance of 1 m, an adversary that uses a transmitting power of 4 W can manipulate the incubator to decrease the actual temperature from the preset $36^\circ C$ to $29^\circ C$, which is close to the room temperature during our experiment.
Adversaries trick the infant incubator to actuate the cooling system by increasing the skin temperature measurement maliciously.
For instance, with an attack distance of 2 m, an adversary can increase the skin temperature measurement by $4.2^\circ C$ (Fig. \ref{fig_incubator}).
Using the same setup as the heating attack, an adversary in the adjacent room can increase the skin temperature measurement by $3.4^\circ C$.

Moderate hypothermia occurs when the auxiliary temperature of an infant drops below $34.9^\circ C$ and severe hypothermia can be caused when it drops below $32^\circ C$ \cite{severe_hypo}. As we demonstrate, the attack can manipulate the infant incubator to decrease the actual temperature to the room temperature such as $29^\circ C$ without triggering any alarm in the incubator system. The compromised incubator system would put the newborn at risks of serious and potentially life-threatening complications such as hypoxia, acidosis, cardiorespiratory and neurological complications, etc. \cite{severe_hypo,mance2008keeping}.

In our experiments, the time  necessary to manipulate the incubator to raise the actual air temperature of the cabin to $38.5^\circ C$ is less than 10 minutes; and it takes about 30 minutes to drop the actual temperature to below $32^\circ C$.
Nurses usually check and record the axillary temperature of newborns at a specific interval. Four hourly is the general recommended interval \cite{nurse2,gardner2011merenstein}. When instability occurs, the interval can be every 30 to 60 minutes \cite{nurse2,gardner2011merenstein}. Adversaries could exploit these intervals to pursue the attack.

\vspace{-7pt}
\subsubsection{Traceable Thermometers with Alarms}
Traceable thermometers that provide highly-accurate temperature measurements are often used to monitor the quality of temperature-sensitive medication such as vaccines, or biological substances \cite{gazmararian2002vaccine,cdc_thermometer}.
They provide reliable temperature data records to assess the quality of substances being monitored and can raise an alarm when the storage temperature is out of a predefined range.
We investigate our attacks on a Thomas traceable thermometer and a Fisherbrand traceable thermometer.
Our experiments show that the integrity of the temperature data recorded by these thermometers can be compromised by attacks. For instance, with an attack distance of 0.5 m and a transmitting power of 4 W, an adversary can increase the temperature measurement of the Fisherbrand thermometer from $26^\circ C$  to $49^\circ C$ or decrease it to $20^\circ C$.
Malicious manipulations of the measurements can result in a recorded temperature data profile inconsistent with the actual quality of the biologic substances being monitored, which could lead to the waste of effective substances or the misuse of ineffective ones that should be discarded. Also, it is possible for adversaries to manipulate the measured temperature to suppress the alarm while the actual temperature is out of the safety range.

\subsection{Laboratory Applications} 


\subsubsection{Biological Laboratory Equipment} Temperature-based systems are widely used in biological laboratory equipment to preserve biological samples or control the temperature during critical experiments.
These equipment are usually well-designed and are expected to control the temperature accurately because an unstable temperature environment can devastate valuable biological samples or bias the outcomes of experiments.
However, in our experiments, we demonstrate how they can be maliciously compromised by adversaries.

\begin{figure}[t]
\centering
 \includegraphics[scale=0.62]{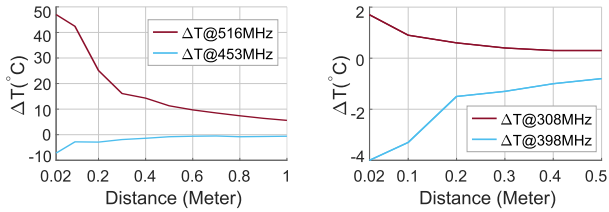}
   \vspace{-3mm}
  \caption{Maximum increase/decrease in the temperature measurement that can be achieved in attack experiments on the hybridization oven (left) and the incufridge (right) with different attack distances. }\label{fig_biology}
   \vspace{-4mm}
\end{figure}

We investigate our attacks on a hybridization oven and an incufridge. The UVP HB-500 hybridization oven accurately controls the temperature of samples in the hybridization process, enabling consistent saturation of sample solutions. It has an internal temperature sensor and is shielded with metal panels, but the gaps between these panels could allow EMI signals to pass through and affect internal circuit components. With an attack distance of 1 m, an adversary can maliciously increase the temperature measurement by $5.6^\circ C$ and trick the hybridization oven to cool down.

The Revolutionary Science RS-IF-202 incufridge can be used to refrigerate or incubate specimens and biological products \cite{incufridger}. The incufridge has an internal temperature sensor and is well-shielded with metal panels. However, we find that EMI signals could enter through the control panel of the device, which can be exploited to spoof its temperature sensor measurement.
In the experiments, we use a transmitting power of 4 W, and we summarize the experimental results in Fig. \ref{fig_biology}.

\vspace{-2pt}
\subsubsection{Thermal Chambers}
Thermal chambers can provide an accurately controlled thermal environment for automatic environmental tests of critical components such as aircraft electronics, satellite antennas, and implantable stents \cite{chamber_app}. Adversarial control or disruptions of these systems could damage expensive components or make the results of environmental tests unreliable.

\begin{figure}
\centering
\includegraphics[scale=0.61]{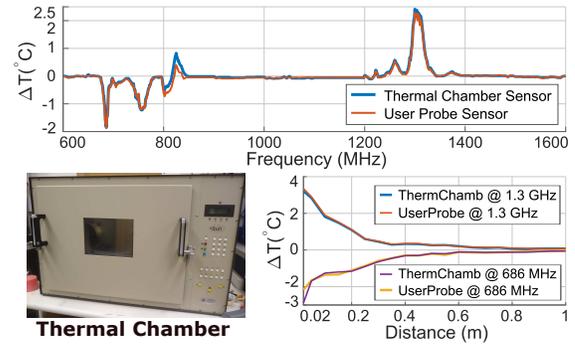}
\vspace{-3mm}
\caption{Temperature offsets induced on the thermal chamber with different attack distances using a transmitting power of 3.2 W. Note that the injection affect both the sensors in similar way despite the chamber shield. }
\label{fig_thermal} \vspace{-3mm}
\end{figure}

We investigate our attacks on a Sun Electronic Systems EC12 thermal chamber that is intended for automated test systems and laboratory applications \cite{ec1tc}.
This well-shielded metal chamber is equipped with two K-type thermocouples: The first one (thermal chamber sensor) is hidden behind the control panel and it measures the internal temperature of the chamber; The second one (the user probe) can be used to directly monitor the temperature of the device under test.
We set and maintain the temperature of the chamber at $30 ^\circ$C, then we turn off the heater circuit breaker to ensure that only the temperature offset caused by the attack is measured.
In our experiments, we point the antenna towards the double-paned glass window of the chamber and sweep a frequency range of 550 MHz to 1.6 GHz using a transmitting power of 35 dBm, which is equivalent to 3.2 W. We monitor the temperature variations in both the thermal chamber probe and the user probe. Although the sensors are placed in different locations of the chamber and the thermal chamber sensor is protected by a metal internal panel, our attack induces similar effects on both of the sensors simultaneously (Fig. \ref{fig_thermal}). We also measure the maximum increases or decreases that can be induced in the temperature measurements with different attack distances.

\subsubsection{3D Printers} In 3D printers, extruders are crucial components that are responsible for heating and expelling the building material (filament).
The temperature control system of an extruder constantly monitors and adjusts the temperature of its heating chamber.
During the building process, the temperature of the heating chamber must be kept within a certain tolerance range to ensure the quality of the build and prevent buildups of the filament \cite{evans2012practical}. Compromising the temperature sensor reliability in this fundamental phase could disrupt the printing process or damage the product quality.
We investigate our attacks on two different extruder models: the MakerBot Smart Extruder and the MakerBot Smart Extruder + (Plus).
We install these extruders onto two identical MakerBot Replicators 3D printers. Both of the two models use K-type shielded thermocouple sensors.
Note that we do not turn on the extruder heating/cooling cycle to prevent damage to the heating chamber. We wait until the temperature of the extruder naturally reaches the equilibrium at room temperature (23$^\circ$C) before starting the attack.

\begin{figure}[t]
\centering
 \includegraphics[scale=0.31]{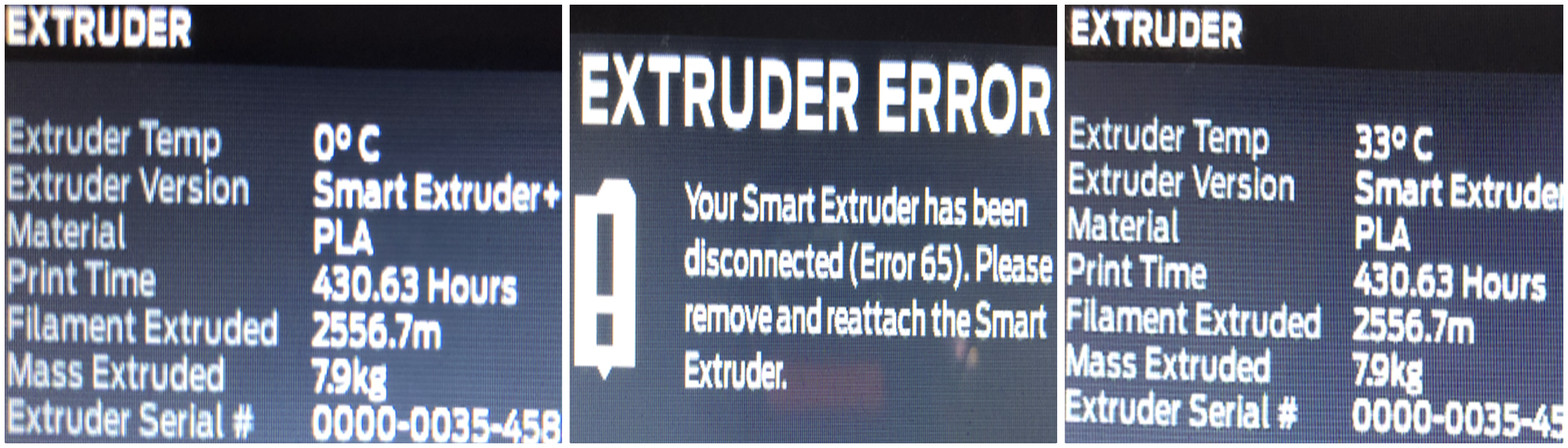}
   \vspace{-4mm}
  \caption{Results of our attack experiments on 3D printers. With a frequency of 400 MHz, the attack causes the disconnection of the extruder (left, middle). With an attack frequency of 1 GHz, the temperature perceived is $10^\circ C$ higher than the actual temperature of $23^\circ C$ (right).}\label{fig:extruderError} \vspace{-4mm}
\end{figure}

We test the frequency range of 400 MHz to 1 GHz, observing the temperature change of the extruder on the 3D printer's display.
During the test, we observe two main effects: 1) With an attack frequency of 400 MHz, the user panels of both of extruder models show that the extruder temperature is zero. Even after reloading the extruder monitoring system, the displayed temperature remains zero (Fig.  \ref{fig:extruderError} left). When we start the ``preheat" functionality, the device displays an extruder disconnection error message (Fig. \ref{fig:extruderError} middle). 2) With an attack frequency of 1 GHz, we can increase the temperature measurement of the Smart Extruder Plus by a maximum of 10 $^\circ$C compared to the baseline temperature (Fig. \ref{fig:extruderError} right).
In this case, the system does not give any error messages or special indications in the user panel when the measured extruder temperature is changed. Therefore, adversaries can spoof the temperature measurement to manipulate the temperature control system in the extruder without being detected by the system. In the experiments, we use a transmitting power of 3.2 W and we are able to induce a temperature change of $0.5^\circ C$ at a maximum attack distance of 25 cm.

\subsection{Consumer PID Controllers}

We study the effect of our attacks on three consumer PID control modules: the Inkbird ITC-100VH, ITC-1000F, ITC-100RH. Although the modules we test are mainly used in IoT applications, devices with similar functions can be found in critical industrial and manufacturing applications \cite{alarm_intro,alarm_intro2,alarm_intro3}. The three modules are equipped with different types of temperature sensors. These devices can be used to limit or maintain the temperature of a target environment in a specific range. When the device detects a temperature that is out of the predefined range, it can raise the alarm to alert users or switch on/off the circuit of a heating or cooling element. Manipulation of temperature measurements can undermine the alarm function even when the actual temperature is out of the predefined range. Our experiments show that these modules are not well-shielded and can be susceptible to adversarial control with a relatively long attack distance. For instance, from a distance of 2 m, the attack can maliciously increase the temperature measurement of the ITC-100RH controller by a maximum of $32.9^\circ C$ with an attack frequency of 453 MHz and a transmitting power of 4 W.

\section{Countermeasures}

Usually manufacturers implement filters to reduce external and internal electromagnetic interference, such as common-mode or differential-mode filters on the amplifier input \cite{emiopamp}. However, as we demonstrate in our work, out-of-band EMI can induce AC signals that bypass generic filtering and be internally rectified through the amplifier input, output, or power supply pins. Although EMI defenses are known and some are already applied to certain critical applications \cite{weston2016electromagnetic}, consumer electronics are less protected against malicious attacks that affect temperature sensors.
In this section, we discuss and simulate several passive and active methods to detect or prevent EMI effects on temperature sensors.
\vspace{-6pt}
\subsection{Hardware Defenses} \vspace{-2pt}
Traditional hardware defenses can take various forms according to the level of mitigation adopted and cost/performance limitations.

\vspace{-2pt}
\paragraph{Shielding}
Designing short shielded wires between the temperature sensors and amplifier inputs is a good practice to avoid long leads acting as antennas and picking up interference. However, the common-mode noise induced by the antenna can become normal mode at the point where the cables are connected to the circuit. This happens because of the difference between the terminal impedance of the cable and the
terminal impedance of receiver circuit \cite{wang2010analysis}. In this case, a mitigation of the attack consists in adding terminating resistors to the contact points. EMI enclosures can also be used to block interference. However, openings in the shield are often required to accommodate switches, connectors, indicators, or to provide  ventilation. These openings may compromise  shielding effectiveness by providing paths for high-frequency interference to enter the circuit board \cite{morrison}. Moreover, it requires a careful thermal modeling of the system \cite{thermalmodeling}.
Another approach consists of sensor shielding when the temperature sensor needs to be externally exposed. In this case, shielding is only effective against interference if it provides a low impedance path to ground. However, some data acquisition systems require the temperature sensor to be grounded, such as thermocouple or RTD probes used in industrial processes \cite{thermocouple}. When both the shield and temperature sensors are grounded, a ground-loop current can appear to the amplifier input terminals due to the difference of potential developed between the sensor ground and the amplifier ground connection \cite{note}. When the EMI induces common mode noise, the interference can pass through the ground of the shield, creating a ground-loop current that can potentially generate the rectification effect. Some techniques can reduce but not eliminate the phenomenon, such as making the shield connection to ground as close as possible to the sensor connection to ground, or using only the ground terminal of the amplifier to connect to the shield without connecting the shield to the amplifier end.

\vspace{-4pt}
\paragraph{Active and Passive Filters}
In the case of op-amps and in-amps, manufacturers apply low-pass filters at the amplifier input pins to reduce the EMI signal energy from the input lines.
In IC temperature sensors that use an inverting op-amp (e.g., LM35), a filter capacitor is placed between equal value resistors, while in IC temperature sensors (e.g., LM335) that use non-inverting op-amp, the filter capacitor is directly connected to the op-amp input. Precision in-amps in RTD and thermocouples sensors use two low-pass filters to suppress common-mode signals in each input lane and one capacitor to suppress differential-mode signals between the two amplifiers input terminals \cite{rfi_rectify_mt96}. These filters are not sufficient for a complete attack mitigation due to the lines asymmetry and frequency range with respect to our injected interference. For example, in thermocouples, the asymmetry between the lines is exacerbated due to the two different conductors tied together. For these reasons, high precision temperature instruments contain additional isolation circuits and active low-pass filters connected to the amplifier input terminals to isolate the field-side and system-side circuitry \cite{techtip}. Another protection method uses a composition of instrumentation amplifiers: three in-amps, two of these correlated to one another and connected in antiphase \cite{kitchin2004designer}.

Choke-based filters can be also used as alternative for in-amp input filtering \cite{kitchin2004designer}. Despite the good noise suppression, the materials used for the inductance cores can heavily affect the filter performance for high frequency EMI, making the system vulnerable to injection attacks \cite{weber2005radio}.

Amplifier outputs also need to be protected from EMI, since the interference injected on an output line couples back into the amplifier input where they are rectified and appear again on the output as a DC offset. An RC filter and/or a ferrite bead in series with the amplifier output are the simplest and inexpensive solutions to reduce the DC offset. However, for temperature systems, the output filtering is often limited to the line frequency and its harmonics (50 Hz/60 Hz) due to the interference noise generated when systems operate from the main power supply \cite{Duff2010TwoWT, mccarthyadc}.
\vspace{-4pt}
\subsection{Software Defenses} \vspace{-2pt}
Many current temperature control systems use multiple sensors to continuously monitor the thermal state of different measurement points, or as multiple temperature reference values \cite{rfi_rectify_mt96, frolik2001confidence}.
In critical infrastructure sectors such as energy and healthcare, redundant sensors are used to generate time-dependent estimates of the critical points \cite{jin2009redundant, ray1991introduction}.
Similar to sensor redundancy, sensor fusion techniques might be used to combine data from different sensors in order to produce the best estimation of the true state of a system and decrease the system's dependence on a single sensor \cite{Ivanov2016AttackResilientSF}. In systems that rely on temperature sensors, literature provides various software countermeasures based on sensor fusion  \cite{kong2005distributed, frameworksensor}.
However, in our experiments we demonstrate how physical proximity causes similar temperature sensors to be affected by similar attack effects (see Fig. \ref{fig_thermal}). In turn, this increases the difficulty for a sensor fusion algorithm to detect the anomalies in small and self-contained systems, such as thermal chambers, or incubators.
In addition, complex sensor fusion techniques require building models of the attacks effects on different sensors, using machine learning-based or statistical techniques to recognize the anomalies~\cite{khaleghi2013multisensor}. Therefore, to cover all the possible attack effects, these approaches require accurate parameter tuning and an exhaustive training phase. This might not be feasible to achieve. Furthermore, if the attack gradually changes the sensor data, or the operating conditions of the system change overtime, the sensor fusion algorithms might not be able to recognize the attack from the normal system behavior.

Other techniques focus on detecting injection attacks at the process level. A process-level intrusion detection system monitors sensors to determine if the physical process drifts from the normal or expected behavior. Common approaches include building Linear Dynamical State-Space (LDS) models of the physical process, or use machine learning and data mining to detect anomalies in the system behavior ~\cite{aoudi2018truth}. Although such approaches might detect anomalous events, models are difficult to build, as they require high effort in simulating and testing all possible attack vectors, and building a complete and highly detailed model of the physical process and interaction is not always feasible. Furthermore, machine learning methods that do not require a model of the physical process involve critical feature extraction and parameter-tuning phases that are often hard to automate and update on the discovery of a new attack vector.
In addition, the systems that implement these kind of techniques need to continuously check if each sensor measurement drifts from the normal behavior captured during the training phase, drastically augmenting the computational and power resource costs.

Sensor redundancy, process-based techniques, and sensor fusion may significantly increase the effort an adversary must make to conduct an attack. However, implementing sophisticated software-based defenses remains arduous in large-scale consumer electronic devices.

\subsection{Hardware Anomaly Detection System} \vspace{-2pt}
For critical applications where it is not possible to implement complete shielding, or an effective mitigation filtering of the system and the sensor(s) - such as incubators - detecting the presence of attack attempts becomes crucial for verifying and maintaining temperature data reliability. A detection circuit can be used as a trigger for emergency measures - such as activating a safe mode where the system restricts its reliance on sensor data. To defend against EMI on cardiac implantable medical devices, Foo Kune et al. \cite{kune2013ghost} proposed a cardiac probe to cross-check whether readings from a cardiac signal coincides with the expected values. Wang et~al. \cite{wang2017sonic} proposed an additional microphone to detect resonating sound that can affect MEMS gyroscopes.
Based on our results, an effective defense for temperature-sensor-based systems that maintains the reliability of the temperature data should account for the frequencies that can induce a rectification effect in the amplifier output signal.
Based on this frequency analysis, manufacturers can modify the design of their system to detect and react to attacks in the frequency bands of EMI signals. We propose a hardware anomaly detector to identify malicious signal and provide feedback about the reliability of the measurement data.

\begin{figure}
\centering
\includegraphics[scale=0.195]{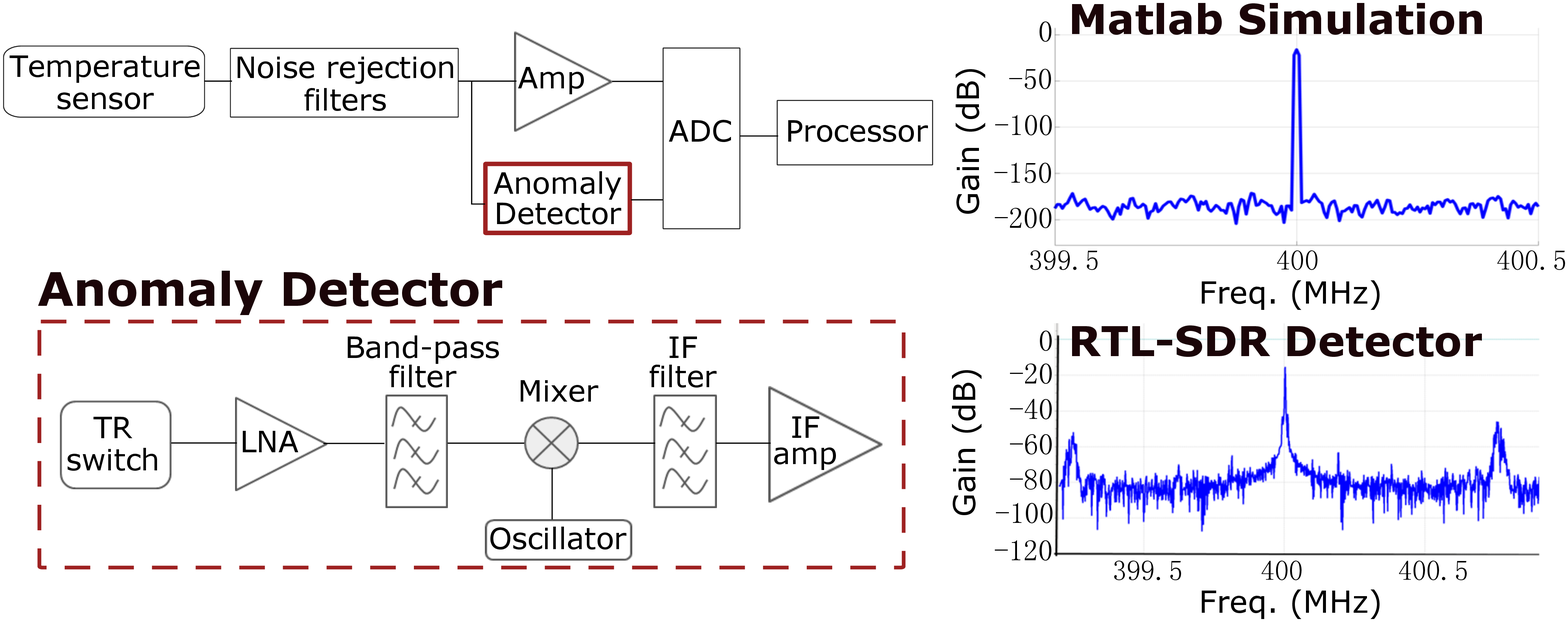}
\vspace{-2mm}
\caption{Block diagram and calculated gain of the anomaly detector based on superheterodyne method.}\label{fig_anomaly}\vspace{-3mm}
\end{figure}

\vspace{-4pt}
\paragraph{Design of the Anomaly Detector}
The EMI signal induced by our attack can appear in many different points close to the amplifier where isolation circuitry and filters don't properly block the high frequency signals.
A detector that can measure these signals can be implemented by connecting a low noise amplifier (LNA) and a band-pass filter to the points (such as a trace or wire) sensitive to the malicious signal (Fig. \ref{fig_anomaly}).
By adopting the superheterodyne technique typical of AM receivers \cite{super}, the EMI frequency bands that cause significant DC offset variations can be down-converted to an intermediate frequency (IF). Down-conversion can be achieved by using a mixer and local oscillator. As a result, the use of this technique allows for a ``tunable filter", which we can utilize for a tunable detector.
Once the signal is digitally converted, amplitude and phase information of the malicious signals at the intermediate frequency can be then analyzed by the processor: (1) providing feedback on the temperature data reliability, (2) allowing the estimation of the measurement error, and (3) compensating it at the software level. The detector can be periodically activated when a temperature measurement is required. A variable oscillator can be used to select multiple vulnerable frequency bands.

\paragraph{Simulation Model and Evaluation}
We simulate the detector against attacks on thermocouple sensors of the same type used in the thermal chamber. In this simulation, our detector can detect signals from 550 MHz to 1 GHz - the range which major affected the sensor (shown in Fig. \ref{fig_ldc4ku}). The simulation was designed using the Simulink environment \cite{simulink}, and consists of an LNA filter with 50 dB gain 3-order Butterworth band-pass filter, followed by a mixer block to down-convert the simulated EMI frequency to an IF frequency of 400 MHz, and an IR filter for filtering the spectral image components. Then, a subsequent 3-order Butterworth IF filter block is followed by an IF amplifier block with 100 dB gain and a noise figure of 2.5 dB. An RF Blockset testbench is used to simulate the EMI injection attack with an emitting power of 35 dBm.


To evaluate our design, we use a Software-Defined Radio (SDR) RTL-SDR device \cite{sdr}. We choose the Realtek RTL2832U chipset with the R820T2 tuner chip that can detect frequencies from 500 kHz up to 1.75 GHz. An RF exposed connection, collocated with the temperature sensor breakout board, is followed by an RF filter and an LNA amplifier at 50 dB. A mixer with a local oscillator is used for the frequency transposition. The detector also uses Automatic Gain Control (AGC), where the gain varies with the available input power level.
As a proof of concept demonstration, we successfully selectively detect a malicious signal at a 3 meter distance from the transmitting antenna, in open air, at a frequency of 503 MHz (corresponding to one of the major effective peaks in Fig. \ref{fig_ldc4ku}). The signal is down-converted to 400 MHz (as shown in Fig. \ref{fig_anomaly}). By varying the local oscillators frequency, the detector can also isolate the other vulnerable frequency bands.

\begin{table*}[t]
	\centering
	\setlength\tabcolsep{3pt} 
	\begin{threeparttable}
		\captionsetup{width=\linewidth}
		\caption{Comparisons between previous studies and our work, including the targeted systems, the affected components, and the effect induced by the attacks.}\label{tab:relatedwork}
		\vspace{-5mm}
		\begin{tabularx}{\linewidth}{|c|c|c c c c|c c c c c c|c|c|}
			\cline{1-12}
			\multicolumn{1}{|c|}{\multirow{2}{*}{\textbf{Paper}}} & \multicolumn{1}{c|}{\multirow{2}{*}{\textbf{System}}}&
			\multicolumn{4}{c|}{\textbf{Exploited Non-linearity Effect}} &
			\multicolumn{6}{c|}{\textbf{Affected Component}} \\  
			\cline{3-12}
			& & Demodulation & Saturation & Aliasing & Rectification & Transd. & Wire & Filter & Amp. & ADC & GPIO  \\
			\cline{1-12}
			\cite{kune2013ghost} & Microphone \bigstrut[t] & \CIRCLE & \Circle &\CIRCLE & \Circle & \CIRCLE & \CIRCLE & \CIRCLE & \CIRCLE & \CIRCLE & \Circle \\
			\cline{2-12}
			& Implantable Medical Dev. \bigstrut[t] & \Circle & \LEFTcircle &\Circle & \Circle & \Circle & \CIRCLE & \CIRCLE & \Circle & \Circle & \Circle \\
			\cline{1-12}
			
			\cite{yan2016can} & Radar \bigstrut[t] & \Circle & \CIRCLE &\Circle & \Circle & \CIRCLE & \Circle & \Circle & \CIRCLE & \Circle & \Circle  \\
			\cline{1-12}
			
			\cite{ware2017effects, selvaraj2018electromagnetic,selvaraj2018intentional} & Microcontroller \bigstrut[t] & \Circle & \Circle &\Circle &\CIRCLE & \Circle & \Circle & \Circle & \Circle & \CIRCLE & \CIRCLE \\
			\cline{1-12}
			
			\cite{delsing2006susceptibility} & Sensor Network \bigstrut[t] & \Circle & \LEFTcircle &\LEFTcircle &\LEFTcircle & \Circle & \CIRCLE & \LEFTcircle & \LEFTcircle & \LEFTcircle & \LEFTcircle \\
			\cline{1-12}
			
			\cite{esteves2018unlocking} & Drone \bigstrut[t] & \Circle & \LEFTcircle &\LEFTcircle &\Circle & \LEFTcircle & \LEFTcircle & \LEFTcircle & \LEFTcircle & \LEFTcircle & \Circle \\
			\cline{1-12}
			
			Our work & Temperature Sensor \bigstrut[t] & \Circle & \Circle & \Circle & \CIRCLE & \Circle & \Circle & \CIRCLE & \CIRCLE & \Circle & \Circle \\
			\cline{1-12}
			
		\end{tabularx}
		\begin{tablenotes}
			\item \hspace{0.2in}\CIRCLE Verified \hspace{0.1in}  \LEFTcircle Uncertain/Unverified \hspace{0.1in} \Circle Not applicable
		\end{tablenotes}
	\end{threeparttable} \vspace{-5mm}
\end{table*}

\section{Related Work}

Analog sensor circuits are especially susceptible to EMI. Various works \cite{giechaskiel2019framework, giechaskiel2019sok, fu2018risks} demonstrate the exploitability of the non-linearities of different circuit components to cause system malfunctions or sensor misreadings (see Table \ref{tab:relatedwork}).

Foo Kune et al. \cite{kune2013ghost} showed that bogus signals can be injected through low-power EMI into analog sensors such as microphones, and implantable medical devices such as pacemakers and defibrillators. Their attack method exploited the generation of sub-harmonics of injected high frequency signals passing through common circuit components (e.g. wires, capacitors, amplifiers, and ADCs). This unintentional demodulation effect down-converts the high frequency signals into low frequency ones. In turn, these low frequency signals are able to pass the protective low-pass filters and enter into the system, compromising its functionality. In automotive field, Yan et al. \cite{yan2016can} intentionally saturated Millimeter-Wave Radar by injecting strong interference, causing sensor denial-of-service in cars.
Unlike these previous works, our EMI injection exploits the rectification effect present in the internal circuitry of operational and instrumentation amplifiers used in temperature-based control systems.

Delsing et al. \cite{delsing2006susceptibility} and Esteves et al. \cite{esteves2018unlocking} empirically observed the general reaction of specific cyber-physical systems under strong interference. Delsing et al. verified the susceptibility of a MULLE node sensor network \cite{johansson2004mulle}. They observed disturbances in the Bluetooth communication, data losses and occasionally rebooting of the sensor network node. They also tested the sensor interface using a temperature sensor, revealing a vulnerability of the device due to the use of a long non-shielded connection between the sensor and the MULLE-device. Esteves et al. investigated a common-off-the-shelf (COTS) civilian quadricopter. They listed several reading errors induced in the drone sensors and interfaces by continuous interference, without exploring the causality of the measured effect.

Recent studies \cite{ware2017effects, selvaraj2018electromagnetic,selvaraj2018intentional} investigated the injection of strong near-field interference to modify the input voltage of GPIO pins in microcontrollers.
In particular, the authors used EMI injection to induce a rectification effect in the Electr-Static Discharge (ESD) protection circuit. The ESD protection circuit is used in microcontrollers GPIO pins to prevent the ADCs to be exposed to out-of-band input voltage when connected to an external analog or a digital sensor. In contrast with these works, our rectification attack directly affects the sensor amplification stage, and in particular the internal transistors in analog sensor amplifiers, before the connection with a microcontroller or the analog-to-digital conversion stage. In addition, instrumentation and operational amplifiers that work with low bias currents such as in temperature sensors, often do not implement external ESD protection circuits at the amplifier input, but only external current limiting resistors \cite{bryant2000protecting}. This approach is used because it provides adequate protection against overvoltage, it does not provoke high current leakage at increasing temperature as it happens using ordinary diodes, and the resistors are already present in the signal conditioning chain, since they constitute part of the low-pass filters used to reject differential and common-mode noise.

\vspace{-3pt}
\paragraph{Physical-level Attacks on Sensors}
Sensors are fundamental for cyber-physical systems such as autonomous vehicles, drones, and medical devices. Existing security studies on sensors have shown how they can be compromised by different kinds of signal injections other than EMI such as mechanical waves (i.e. sound), and light.
For instance, by injecting different types of light signal using lasers, Park et al. \cite{park2016ain} compromised medical infusion pumps to make them over/under-infuse, while Petit et al. \cite{Petit2015RemoteAO} and Shin et al. \cite{shin2017illusion} generate fake obstacles in LiDARs systems for automotive applications.
Other works demonstrate how intense acoustic waves can incapacitate or manipulate some models of micro-electro-mechanical systems (MEMS) inertial sensors \cite{son2015rocking,trippel2017walnut,tu18injected,wang2017sonic}, while Zhang et al. \cite{zhang2017dolphinattack} used ultrasonic waves to send inaudible commands to voice assistance systems, such as Google Home and Alexa.
Similar to our work, these attacks modulate the malicious signal on top of a carrier to infiltrate the system. However, they exploit different physical phenomena, such as the demodulation effect, or aliasing, rather than rectification. For this reason, defenses that mitigate these effects might be not sufficient to mitigate rectification attacks, since the physics principle exploited is different.

The novelty of our work stands on this new attack vector not yet explored by previous research on sensor attacks. Further, we show how this vulnerability of amplifiers can affect different analog temperature sensors that use similar signal conditioning process.

\vspace{-2pt}

\section{Discussions}
\vspace{-0.5pt}

\subsection{Limitations}
\vspace{-0.5pt}
In our study we only consider commercial temperature-based systems that use analog temperature sensors. Our analysis focuses on low-power attacks (less than 4 W) in the Ultra High Frequency range (UHF) 300 MHz - 3 GHz. These assumptions are acceptable considering that our work shows how the rectification attack can be successful with a low-power injection, even if the target system already employs traditional EMI defenses. Also, we assume that an adversary can attempt the attack with little effort by building a small device or modifying a commercial system (e.g. a walkie-talkie) that can emit EMI signals in the vulnerable frequency range. Although the attack distance can be increased with specialized equipment and higher transmitting power, our goal is to demonstrate that simple amplitude-modulated EMI can induce a controllable voltage offset in temperature sensing circuits large enough to deceive and manipulate a target system.

To improve the attack success rate, an adversary might need some additional information regarding the target device, such as the presence of automatic temperature alarms and their threshold values. These information can be retrieved from the publicly available manuals and datasheets of the target system.

During our experiments, we observe that the amount of induced DC offset can be affected by various factors, including the noise rejection circuitry and shield used in the target system, the characteristics of the antenna used to perform the attack (e.g., directional, monopole, etc.) and its orientation with respect to the target device. In addition, to optimize the attack effect, the adversary often needs to position the antenna to target the parts of the victim system that are usually more susceptible (e.g., the temperature sensor transducer, the control panel, etc.).

\vspace{-8pt}
\subsection{Attack Generalization} \vspace{-2pt}

By exploiting this hardware-level vulnerability, adversaries could also affect systems equipped with different classes of sensors that use similar signal conditioning processes. For example, we find that pressure or pH sensors may also be susceptible to adversarial control through rectification attacks, since the transducer signal of these sensors is weak and requires an amplification stage similar to temperature sensors.

\paragraph{Pressure Sensor}
Scales use pressure sensors to measure the weight of an object. Sensor wires distributed inside of the device can make it vulnerable to EMI injection.
We test a CGOLDENWALL high-precision lab digital scale that has an accuracy of 0.01 g, which can be used in jewelry, laboratory measurements.
We are able to decrease the reading of the scale by 6.37 g at a distance of 0.5 m with an attack frequency of 685 MHz.
We also test an Escali L600 L-Series High Precision Lab Scale. At an attack distance of 0.5 m, we can decrease the reading of the scale by 7 g, or increase the reading by 13.9 g.
Using the same attack technique we show in this work, adversaries might be able to spoof the pressure or force measurement in data acquisition or control systems to cause unexpected consequences.

\vspace{-2pt}
\paragraph{PH Sensor}
A pH meter measures a low-level difference in electrical potential between a pH electrode and a reference electrode. We test an Apera Instruments PH700 Benchtop Lab pH Meter that has an accuracy of 0.01 pH.
At an attack distance of 0.5 m, we can increase the measured pH value by 0.42 with EMI signal injections at a frequency of 515 MHz. PH sensors can be used in closed-loop control in SCADA systems such as water treatment facilities. Adversaries might attempt to manipulate the actual pH value to damage the facilities of such systems by exploiting pH sensors.

\section{Conclusion}

Temperature-based control systems fundamentally rely on sensors to make critical decisions. So it is important to assess and improve the resilience of the system in situations when temperature sensors could be compromised. This work showed how adversaries can manipulate these systems to cause unexpected consequences, without tampering with the victim system or triggering temperature alarms. The attack leveraged an unintended rectification effect in amplifiers to control the DC voltage of temperature sensor signals. We validated the attack on sensors and investigated the threat on several real-world temperature control systems.
Our experimental results showed that these systems blindly trust spoofed temperature sensor readings, leading to manipulated decision makings of a victim system.
To mitigate the risks, we discussed several conventional defensive techniques, and proposed a prototype design of an analog anomaly detector to ensure the integrity of temperature sensor signals.

\paragraph{Acknowledgments}
We are in the process to coordinate with ICS-CERT to notify manufacturer companies whose sensors and devices we tested. This work is supported in part by US NSF under grants CNS-1812553 and CNS-1330142.
\vspace{-3mm}

\bibliographystyle{ACM-Reference-Format}
\bibliography{thermo}


\begin{thebibliography}{89}


\ifx \showCODEN    \undefined \def \showCODEN     #1{\unskip}     \fi
\ifx \showDOI      \undefined \def \showDOI       #1{#1}\fi
\ifx \showISBNx    \undefined \def \showISBNx     #1{\unskip}     \fi
\ifx \showISBNxiii \undefined \def \showISBNxiii  #1{\unskip}     \fi
\ifx \showISSN     \undefined \def \showISSN      #1{\unskip}     \fi
\ifx \showLCCN     \undefined \def \showLCCN      #1{\unskip}     \fi
\ifx \shownote     \undefined \def \shownote      #1{#1}          \fi
\ifx \showarticletitle \undefined \def \showarticletitle #1{#1}   \fi
\ifx \showURL      \undefined \def \showURL       {\relax}        \fi
\providecommand\bibfield[2]{#2}
\providecommand\bibinfo[2]{#2}
\providecommand\natexlab[1]{#1}
\providecommand\showeprint[2][]{arXiv:#2}

\bibitem[\protect\citeauthoryear{??}{who}{1997}]%
        {who_thermoregulation}
 \bibinfo{year}{1997}\natexlab{}.
\newblock \bibinfo{title}{{World Health Organization, Maternal and Newborn
  Health/Safe Motherhood. Thermal protection of the newborn: a practical
  guide}}.
\newblock
  \bibinfo{howpublished}{\url{https://apps.who.int/iris/bitstream/handle/10665/63986/WHO_RHT_MSM_97.2.pdf}}.
\newblock


\bibitem[\protect\citeauthoryear{??}{inc}{2001}]%
        {incubator3}
 \bibinfo{year}{2001}\natexlab{}.
\newblock \bibinfo{title}{{Franks Hospital. Babytherm 8000 WB Warming bed.
  Instructions for Use. Page: 18}}.
\newblock
  \bibinfo{howpublished}{\url{http://www.frankshospitalworkshop.com/equipment/documents/infant_incubators/user_manuals/Ginevri
  OGB PolyCare 3 Incubator - User manual.pdf}}.
\newblock


\bibitem[\protect\citeauthoryear{??}{rfi}{2009}]%
        {rfi_rectify_mt96}
 \bibinfo{year}{2009}\natexlab{}.
\newblock \bibinfo{title}{{Analog Devices. RFI Rectification Concepts}.}
\newblock
  \bibinfo{howpublished}{\url{https://www.analog.com/media/en/training-seminars/tutorials/MT-096.pdf}}.
\newblock


\bibitem[\protect\citeauthoryear{??}{inc}{2009}]%
        {incufridger}
 \bibinfo{year}{2009}\natexlab{}.
\newblock \bibinfo{title}{{Revolutionary Science. Product descriptions of the
  RS-IF-202 Incufridge.}}
\newblock
  \bibinfo{howpublished}{\url{https://wikisites.mcgill.ca/djgroup/images/4/41/Incufridge_19L_Model_RS-IF-202.pdf}}.
\newblock


\bibitem[\protect\citeauthoryear{??}{ec1}{2011}]%
        {ec1tc}
 \bibinfo{year}{2011}\natexlab{}.
\newblock \bibinfo{title}{{Sun Electronic Systems. Model EC1X environmental
  chamber user and repair manual}}.
\newblock
  \bibinfo{howpublished}{\url{http://eecs.oregonstate.edu/matdev/man/Sun_Electronic_Systems_Environmental_Chamber_EC1X.PDF}}.
\newblock


\bibitem[\protect\citeauthoryear{??}{who}{2011a}]%
        {who_infantdeath}
 \bibinfo{year}{2011}\natexlab{a}.
\newblock \bibinfo{title}{{World Health Organisation (WHO). Core Medical
  Equipment: Incubator, Infant (Page 31)}}.
\newblock
  \bibinfo{howpublished}{\url{http://apps.who.int/medicinedocs/documents/s22062en/s22062en.pdf}}.
\newblock


\bibitem[\protect\citeauthoryear{??}{who}{2011b}]%
        {who_bloodcoldchain}
 \bibinfo{year}{2011}\natexlab{b}.
\newblock \bibinfo{title}{{World Health Organisation (WHO)}. The blood cold
  chain.}
\newblock
  \bibinfo{howpublished}{\url{https://www.who.int/bloodsafety/processing/who_eht_11_04_en.pdf}}.
\newblock


\bibitem[\protect\citeauthoryear{??}{sev}{2013}]%
        {severe_hypo}
 \bibinfo{year}{2013}\natexlab{}.
\newblock \bibinfo{title}{Champlain Maternal Newborn Regional Program (CMNRP).
  Newborn Thermoregulation Self-Learning Module}.
\newblock
  \bibinfo{howpublished}{\url{http://www.cmnrp.ca/uploads/documents/Newborn_Thermoregulation_SLM_2013_06.pdf}}.
\newblock


\bibitem[\protect\citeauthoryear{??}{spa}{2015}]%
        {spark_31855}
 \bibinfo{year}{2015}\natexlab{}.
\newblock \bibinfo{title}{{Sparkfun. Schematics of the SparkFun MAX31855K
  Thermocouple Breakout}.}
\newblock
  \bibinfo{howpublished}{\url{https://cdn.sparkfun.com/datasheets/Sensors/Temp/SparkFun_Thermocouple_Breakout_v10.pdf}}.
\newblock


\bibitem[\protect\citeauthoryear{??}{who}{2015}]%
        {who_coldchain01}
 \bibinfo{year}{2015}\natexlab{}.
\newblock \bibinfo{title}{{World Health Organisation (WHO)}. Vaccine management
  handbook: How to monitor temperatures in the vaccine supply chain}.
\newblock
  \bibinfo{howpublished}{\url{https://apps.who.int/iris/bitstream/handle/10665/183583/WHO_IVB_15.04_eng.pdf}}.
\newblock


\bibitem[\protect\citeauthoryear{??}{ada}{2018}]%
        {ada_8495}
 \bibinfo{year}{2018}\natexlab{}.
\newblock \bibinfo{title}{{Adafruit. Analog Output K-Type Thermocouple
  Amplifier - AD8495 Breakout}.}
\newblock \bibinfo{howpublished}{\url{https://www.adafruit.com/product/1778}}.
\newblock


\bibitem[\protect\citeauthoryear{??}{vac}{2018}]%
        {vaccine_cdc}
 \bibinfo{year}{2018}\natexlab{}.
\newblock \bibinfo{title}{{Centers for Disease Control and Prevention (CDC)}.
  Vaccine Storage and Handling}.
\newblock
  \bibinfo{howpublished}{\url{https://www.cdc.gov/vaccines/pubs/pinkbook/vac-storage.html}}.
\newblock


\bibitem[\protect\citeauthoryear{??}{inc}{2018a}]%
        {incubator1}
 \bibinfo{year}{2018}\natexlab{a}.
\newblock \bibinfo{title}{{Franks Hospital. Air Shields Isolette C-100, C-200
  Infant Incubator Service manual. Section:7}}.
\newblock
  \bibinfo{howpublished}{\url{http://www.frankshospitalworkshop.com/equipment/documents/infant_incubators/service_manuals/Air-Shields_Isolette_C-100,C-200_Infant_Incubator_-_Service_manual.pdf}}.
\newblock


\bibitem[\protect\citeauthoryear{??}{inc}{2018b}]%
        {incubator2}
 \bibinfo{year}{2018}\natexlab{b}.
\newblock \bibinfo{title}{{International Biomedical. AirBorne 185A+ Transport
  Incubator Service Manual}}.
\newblock
  \bibinfo{howpublished}{\url{https://www.int-bio.com/wp-content/uploads/2016/06/185A-Service-Manual-English-Rev-C.pdf}}.
\newblock


\bibitem[\protect\citeauthoryear{??}{tec}{2018}]%
        {techtip}
 \bibinfo{year}{2018}\natexlab{}.
\newblock \bibinfo{title}{{IOtech. Grounding and Shielding Considerations for
  Thermocouples, Strain Gages, and Low-Level Circuits}.}
\newblock
  \bibinfo{howpublished}{\url{http://www.mccdaq.com/pdfs/techtip/techtip_60201.pdf}}.
\newblock


\bibitem[\protect\citeauthoryear{??}{sim}{2018}]%
        {simulink}
 \bibinfo{year}{2018}\natexlab{}.
\newblock \bibinfo{title}{{Mathworks. Superheterodyne Receiver Using RF Budget
  Analyzer App}}.
\newblock
  \bibinfo{howpublished}{\url{https://www.mathworks.com/help/rf/examples/superheterodyne-receiver-using-rf-budget-analyzer-app.html}}.
\newblock


\bibitem[\protect\citeauthoryear{??}{set}{2018a}]%
        {setup_amp}
 \bibinfo{year}{2018}\natexlab{a}.
\newblock \bibinfo{title}{{Minicircuits ZHL-4240W broad-band amplifier}.}
\newblock
  \bibinfo{howpublished}{\url{https://www.minicircuits.com/pdfs/ZHL-4240W.pdf}}.
\newblock


\bibitem[\protect\citeauthoryear{??}{rtd}{2018}]%
        {rtdomega}
 \bibinfo{year}{2018}\natexlab{}.
\newblock \bibinfo{title}{{Omega Co. Introduction to Resistance Temperature
  Detectors}}.
\newblock
  \bibinfo{howpublished}{\url{https://www.omega.com/prodinfo/rtd.html}}.
\newblock


\bibitem[\protect\citeauthoryear{??}{set}{2018b}]%
        {setup_ant}
 \bibinfo{year}{2018}\natexlab{b}.
\newblock \bibinfo{title}{{RFSpace LPDAMAX Wide-band PCB Log Periodic Antenna
  }.}
\newblock
  \bibinfo{howpublished}{\url{http://rfspace.com/RFSPACE/Antennas_files/LPDA-MAX.pdf}}.
\newblock


\bibitem[\protect\citeauthoryear{??}{ala}{2018a}]%
        {alarm_intro2}
 \bibinfo{year}{2018}\natexlab{a}.
\newblock \bibinfo{title}{Southwest Heater \& Controls. Limit and Alarm
  Controllers}.
\newblock
  \bibinfo{howpublished}{\url{https://www.swhc.com/limit-alarm-controller.htm}}.
\newblock


\bibitem[\protect\citeauthoryear{??}{ala}{2018b}]%
        {alarm_intro}
 \bibinfo{year}{2018}\natexlab{b}.
\newblock \bibinfo{title}{Temperature Controller Basics Handbook}.
\newblock
  \bibinfo{howpublished}{\url{https://www.instrumart.com/pages/283/temperature-controller-basics-handbook}}.
\newblock


\bibitem[\protect\citeauthoryear{??}{ala}{2018c}]%
        {alarm_intro3}
 \bibinfo{year}{2018}\natexlab{c}.
\newblock \bibinfo{title}{West Control Solutions Co. Temperature Monitors and
  Limiters}.
\newblock
  \bibinfo{howpublished}{\url{https://www.west-cs.com/assets/Brochures/BR-LD-2-US-1903-web-1.0.pdf}}.
\newblock


\bibitem[\protect\citeauthoryear{??}{ada}{2019}]%
        {ada_31865}
 \bibinfo{year}{2019}\natexlab{}.
\newblock \bibinfo{title}{{Adafruit. Digital Output PT100 RTD Temperature
  Sensor Amplifier - MAX31865 Breakout}.}
\newblock \bibinfo{howpublished}{\url{https://www.adafruit.com/product/3328}}.
\newblock


\bibitem[\protect\citeauthoryear{??}{cdc}{2019}]%
        {cdc_thermometer}
 \bibinfo{year}{2019}\natexlab{}.
\newblock \bibinfo{title}{{Centers for Disease Control and Prevention (CDC).
  Vaccine Storage and Handling Toolkit}}.
\newblock
  \bibinfo{howpublished}{\url{https://www.cdc.gov/vaccines/hcp/admin/storage/toolkit/storage-handling-toolkit.pdf}}.
\newblock


\bibitem[\protect\citeauthoryear{??}{nur}{2019}]%
        {nurse2}
 \bibinfo{year}{2019}\natexlab{}.
\newblock \bibinfo{title}{Great Ormond Street Hospital for Children - Clinical
  Guidelines: Thermoregulation for neonates}.
\newblock
  \bibinfo{howpublished}{\url{https://www.gosh.nhs.uk/health-professionals/clinical-guidelines/thermoregulation-neonates}}.
\newblock


\bibitem[\protect\citeauthoryear{??}{cha}{2019}]%
        {chamber_app}
 \bibinfo{year}{2019}\natexlab{}.
\newblock \bibinfo{title}{{Thermal Chamber Applications }}.
\newblock
  \bibinfo{howpublished}{\url{https://www.intestthermal.com/products/chambers/applications}}.
\newblock


\bibitem[\protect\citeauthoryear{??}{who}{2019}]%
        {who_bloodcoldchain02}
 \bibinfo{year}{2019}\natexlab{}.
\newblock \bibinfo{title}{{World Health Organisation (WHO)}. Blood cold chain.}
\newblock
  \bibinfo{howpublished}{\url{https://www.who.int/bloodsafety/processing/cold_chain/en/}}.
\newblock


\bibitem[\protect\citeauthoryear{Antonucci, Porcella, and Fanos}{Antonucci
  et~al\mbox{.}}{2009}]%
        {antonucci2009infant}
\bibfield{author}{\bibinfo{person}{Roberto Antonucci},
  \bibinfo{person}{Annalisa Porcella}, {and} \bibinfo{person}{Vassilios
  Fanos}.} \bibinfo{year}{2009}\natexlab{}.
\newblock \showarticletitle{The infant incubator in the neonatal intensive care
  unit: unresolved issues and future developments}.
\newblock \bibinfo{journal}{\emph{Journal of perinatal medicine}}
  \bibinfo{volume}{37}, \bibinfo{number}{6} (\bibinfo{year}{2009}),
  \bibinfo{pages}{587--598}.
\newblock


\bibitem[\protect\citeauthoryear{Aoudi, Iturbe, and Almgren}{Aoudi
  et~al\mbox{.}}{2018}]%
        {aoudi2018truth}
\bibfield{author}{\bibinfo{person}{Wissam Aoudi}, \bibinfo{person}{Mikel
  Iturbe}, {and} \bibinfo{person}{Magnus Almgren}.}
  \bibinfo{year}{2018}\natexlab{}.
\newblock \showarticletitle{Truth will out: Departure-based process-level
  detection of stealthy attacks on control systems}. In
  \bibinfo{booktitle}{\emph{Proceedings of the 2018 ACM SIGSAC Conference on
  Computer and Communications Security}}. ACM, \bibinfo{pages}{817--831}.
\newblock


\bibitem[\protect\citeauthoryear{Bell}{Bell}{2006}]%
        {bell2006iowa_servocontrol}
\bibfield{author}{\bibinfo{person}{Edward~F Bell}.}
  \bibinfo{year}{2006}\natexlab{}.
\newblock \showarticletitle{Servocontrol: Incubator and radiant warmer}.
\newblock \bibinfo{journal}{\emph{Iowa Neonatology Handbook}}
  (\bibinfo{year}{2006}).
\newblock


\bibitem[\protect\citeauthoryear{Benedict and Russo}{Benedict and
  Russo}{1972}]%
        {note}
\bibfield{author}{\bibinfo{person}{Robert~P Benedict} {and} \bibinfo{person}{RJ
  Russo}.} \bibinfo{year}{1972}\natexlab{}.
\newblock \showarticletitle{A note on grounded thermocouple circuits}.
\newblock \bibinfo{journal}{\emph{Journal of Basic Engineering}}
  \bibinfo{volume}{94}, \bibinfo{number}{2} (\bibinfo{year}{1972}),
  \bibinfo{pages}{377--380}.
\newblock


\bibitem[\protect\citeauthoryear{Bryant}{Bryant}{2000}]%
        {bryant2000protecting}
\bibfield{author}{\bibinfo{person}{James Bryant}.}
  \bibinfo{year}{2000}\natexlab{}.
\newblock \showarticletitle{Protecting Instrumentation Amplifiers-All data
  acquisition board designs have to contend with ESO, EMI, and overvoltages.
  Can one solution protect the circuitry against all three hazards?}
\newblock \bibinfo{journal}{\emph{Sensors-the Journal of Applied Sensing
  Technology}} \bibinfo{volume}{17}, \bibinfo{number}{4}
  (\bibinfo{year}{2000}), \bibinfo{pages}{62--69}.
\newblock


\bibitem[\protect\citeauthoryear{David}{David}{2017}]%
        {ware2017effects}
\bibfield{author}{\bibinfo{person}{A.~Ware David}.}
  \bibinfo{year}{2017}\natexlab{}.
\newblock \emph{\bibinfo{title}{Effects of Intentional Electromagnetic
  Interference on Analog to Digital Converter Measurements of Sensor Outputs
  and General Purpose Input Output Pins}}.
\newblock \bibinfo{thesistype}{Ph.D. Dissertation}. \bibinfo{school}{Utah State
  University}.
\newblock


\bibitem[\protect\citeauthoryear{D{\'e}cima, St{\'e}phan-Blanchard,
  L{\'e}k{\'e}, D{\'e}grugilliers, Delanaud, Libert, and Tourneux}{D{\'e}cima
  et~al\mbox{.}}{2013}]%
        {decima2013does}
\bibfield{author}{\bibinfo{person}{Pauline D{\'e}cima}, \bibinfo{person}{Erwan
  St{\'e}phan-Blanchard}, \bibinfo{person}{Andr{\'e} L{\'e}k{\'e}},
  \bibinfo{person}{Loic D{\'e}grugilliers}, \bibinfo{person}{St{\'e}phane
  Delanaud}, \bibinfo{person}{Jean-Pierre Libert}, {and}
  \bibinfo{person}{Pierre Tourneux}.} \bibinfo{year}{2013}\natexlab{}.
\newblock \showarticletitle{Does the incubator control mode influence outcomes
  of low-birth-weight neonates during the first days of life and at hospital
  discharge?}
\newblock \bibinfo{journal}{\emph{Health}} \bibinfo{volume}{5},
  \bibinfo{number}{08} (\bibinfo{year}{2013}), \bibinfo{pages}{6}.
\newblock


\bibitem[\protect\citeauthoryear{Delsing, Ekman, Johansson, Sundberg,
  B{\"a}ckstr{\"o}m, and Nilsson}{Delsing et~al\mbox{.}}{2006}]%
        {delsing2006susceptibility}
\bibfield{author}{\bibinfo{person}{Jerker Delsing}, \bibinfo{person}{Jonas
  Ekman}, \bibinfo{person}{Jonny Johansson}, \bibinfo{person}{Sofia Sundberg},
  \bibinfo{person}{Mats B{\"a}ckstr{\"o}m}, {and} \bibinfo{person}{T Nilsson}.}
  \bibinfo{year}{2006}\natexlab{}.
\newblock \showarticletitle{Susceptibility of sensor networks to intentional
  electromagnetic interference}. In \bibinfo{booktitle}{\emph{International
  Z{\"u}rich Symposium on Electromagnetic Compatibility}}.
\newblock


\bibitem[\protect\citeauthoryear{Duff and Towey}{Duff and Towey}{2010}]%
        {Duff2010TwoWT}
\bibfield{author}{\bibinfo{person}{Matthew~Lawrence Duff} {and}
  \bibinfo{person}{Joseph Towey}.} \bibinfo{year}{2010}\natexlab{}.
\newblock \showarticletitle{Two Ways to Measure Temperature Using Thermocouples
  Feature Simplicity, Accuracy, and Flexibility}.
\newblock \bibinfo{journal}{\emph{A forum for the exchange of circuits,
  systems, and software for real-world signal processing}}
  (\bibinfo{year}{2010}).
\newblock


\bibitem[\protect\citeauthoryear{Esteves, Cottais, and Kasmi}{Esteves
  et~al\mbox{.}}{2018}]%
        {esteves2018unlocking}
\bibfield{author}{\bibinfo{person}{Jos{\'e}~Lopes Esteves},
  \bibinfo{person}{Emmanuel Cottais}, {and} \bibinfo{person}{Chaouki Kasmi}.}
  \bibinfo{year}{2018}\natexlab{}.
\newblock \showarticletitle{Unlocking the Access to the Effects Induced by IEMI
  on a Civilian UAV}. In \bibinfo{booktitle}{\emph{2018 International Symposium
  on Electromagnetic Compatibility (EMC EUROPE)}}. IEEE,
  \bibinfo{pages}{48--52}.
\newblock


\bibitem[\protect\citeauthoryear{Evans}{Evans}{2012}]%
        {evans2012practical}
\bibfield{author}{\bibinfo{person}{Brian Evans}.}
  \bibinfo{year}{2012}\natexlab{}.
\newblock \bibinfo{booktitle}{\emph{Practical 3D printers: The science and art
  of 3D printing}}.
\newblock \bibinfo{publisher}{Apress}.
\newblock


\bibitem[\protect\citeauthoryear{Feteira}{Feteira}{2009}]%
        {feteira2009negative}
\bibfield{author}{\bibinfo{person}{Antonio Feteira}.}
  \bibinfo{year}{2009}\natexlab{}.
\newblock \showarticletitle{Negative temperature coefficient resistance (NTCR)
  ceramic thermistors: an industrial perspective}.
\newblock \bibinfo{journal}{\emph{Journal of the American Ceramic Society}}
  \bibinfo{volume}{92}, \bibinfo{number}{5} (\bibinfo{year}{2009}),
  \bibinfo{pages}{967--983}.
\newblock


\bibitem[\protect\citeauthoryear{Fiori}{Fiori}{2015}]%
        {fiori2015analog}
\bibfield{author}{\bibinfo{person}{Franco Fiori}.}
  \bibinfo{year}{2015}\natexlab{}.
\newblock \showarticletitle{{An analog front end based on chopped signals
  highly immune to RFI}}. In \bibinfo{booktitle}{\emph{Electromagnetic
  Compatibility (APEMC), 2015 Asia-Pacific Symposium on}}. IEEE,
  \bibinfo{pages}{98--101}.
\newblock


\bibitem[\protect\citeauthoryear{Fiori}{Fiori}{2016}]%
        {fiori2016senosr_sign}
\bibfield{author}{\bibinfo{person}{F. Fiori}.} \bibinfo{year}{2016}\natexlab{}.
\newblock \showarticletitle{A Sensor Signal Amplifier Resilient to EMI}.
\newblock \bibinfo{journal}{\emph{IEEE Sensors Journal}} \bibinfo{volume}{16},
  \bibinfo{number}{18} (\bibinfo{year}{2016}), \bibinfo{pages}{7008--7015}.
\newblock


\bibitem[\protect\citeauthoryear{Frolik, Abdelrahman, and Kandasamy}{Frolik
  et~al\mbox{.}}{2001}]%
        {frolik2001confidence}
\bibfield{author}{\bibinfo{person}{Jeff Frolik}, \bibinfo{person}{Mohamed
  Abdelrahman}, {and} \bibinfo{person}{Param Kandasamy}.}
  \bibinfo{year}{2001}\natexlab{}.
\newblock \showarticletitle{A confidence-based approach to the self-validation,
  fusion and reconstruction of quasi-redundant sensor data}.
\newblock \bibinfo{journal}{\emph{IEEE Transactions on Instrumentation and
  Measurement}} \bibinfo{volume}{50}, \bibinfo{number}{6}
  (\bibinfo{year}{2001}), \bibinfo{pages}{1761--1769}.
\newblock


\bibitem[\protect\citeauthoryear{Fu and Xu}{Fu and Xu}{2018}]%
        {fu2018risks}
\bibfield{author}{\bibinfo{person}{Kevin Fu} {and} \bibinfo{person}{Wenyuan
  Xu}.} \bibinfo{year}{2018}\natexlab{}.
\newblock \showarticletitle{Risks of trusting the physics of sensors}.
\newblock \bibinfo{journal}{\emph{Commun. ACM}} \bibinfo{volume}{61},
  \bibinfo{number}{2} (\bibinfo{year}{2018}), \bibinfo{pages}{20--23}.
\newblock


\bibitem[\protect\citeauthoryear{Gaboian}{Gaboian}{2000}]%
        {common}
\bibfield{author}{\bibinfo{person}{Jerry Gaboian}.}
  \bibinfo{year}{2000}\natexlab{}.
\newblock \showarticletitle{A statistical survey of common-mode noise}.
\newblock \bibinfo{journal}{\emph{Analog Applications}} (\bibinfo{year}{2000}).
\newblock


\bibitem[\protect\citeauthoryear{Gardner, Carter, Enzman-Hines, and
  Hernandez}{Gardner et~al\mbox{.}}{2011}]%
        {gardner2011merenstein}
\bibfield{author}{\bibinfo{person}{Sandra~Lee Gardner},
  \bibinfo{person}{Brian~S Carter}, \bibinfo{person}{Mary~I Enzman-Hines},
  {and} \bibinfo{person}{Jacinto~A Hernandez}.}
  \bibinfo{year}{2011}\natexlab{}.
\newblock \bibinfo{booktitle}{\emph{handbook of neonatal intensive care}}.
\newblock \bibinfo{publisher}{St Louis: Mosby Elsevier}. 117--123 pages.
\newblock


\bibitem[\protect\citeauthoryear{Gazmararian, Oster, Green, Schuessler, Howell,
  Davis, Krovisky, and Warburton}{Gazmararian et~al\mbox{.}}{2002}]%
        {gazmararian2002vaccine}
\bibfield{author}{\bibinfo{person}{Julie~A Gazmararian},
  \bibinfo{person}{Natalia~V Oster}, \bibinfo{person}{Diane~C Green},
  \bibinfo{person}{Linda Schuessler}, \bibinfo{person}{Kelly Howell},
  \bibinfo{person}{Janona Davis}, \bibinfo{person}{Marybeth Krovisky}, {and}
  \bibinfo{person}{Samuel~W Warburton}.} \bibinfo{year}{2002}\natexlab{}.
\newblock \showarticletitle{Vaccine storage practices in primary care physician
  offices: assessment and intervention}.
\newblock \bibinfo{journal}{\emph{American journal of preventive medicine}}
  \bibinfo{volume}{23}, \bibinfo{number}{4} (\bibinfo{year}{2002}),
  \bibinfo{pages}{246--253}.
\newblock


\bibitem[\protect\citeauthoryear{Giechaskiel and Rasmussen}{Giechaskiel and
  Rasmussen}{2019}]%
        {giechaskiel2019sok}
\bibfield{author}{\bibinfo{person}{Ilias Giechaskiel} {and}
  \bibinfo{person}{Kasper~Bonne Rasmussen}.} \bibinfo{year}{2019}\natexlab{}.
\newblock \showarticletitle{SoK: Taxonomy and Challenges of Out-of-Band Signal
  Injection Attacks and Defenses}.
\newblock \bibinfo{journal}{\emph{arXiv preprint arXiv:1901.06935}}
  (\bibinfo{year}{2019}).
\newblock


\bibitem[\protect\citeauthoryear{Giechaskiel, Zhang, and Rasmussen}{Giechaskiel
  et~al\mbox{.}}{2019}]%
        {giechaskiel2019framework}
\bibfield{author}{\bibinfo{person}{Ilias Giechaskiel}, \bibinfo{person}{Youqian
  Zhang}, {and} \bibinfo{person}{Kasper~B Rasmussen}.}
  \bibinfo{year}{2019}\natexlab{}.
\newblock \showarticletitle{A Framework for Evaluating Security in the Presence
  of Signal Injection Attacks}.
\newblock \bibinfo{journal}{\emph{arXiv preprint arXiv:1901.03675}}
  (\bibinfo{year}{2019}).
\newblock


\bibitem[\protect\citeauthoryear{Ivanov, Pajic, and Lee}{Ivanov
  et~al\mbox{.}}{2016}]%
        {Ivanov2016AttackResilientSF}
\bibfield{author}{\bibinfo{person}{Radoslav Ivanov}, \bibinfo{person}{Miroslav
  Pajic}, {and} \bibinfo{person}{Insup Lee}.} \bibinfo{year}{2016}\natexlab{}.
\newblock \showarticletitle{Attack-Resilient Sensor Fusion for Safety-Critical
  Cyber-Physical Systems}.
\newblock \bibinfo{journal}{\emph{ACM Trans. Embedded Comput. Syst.}}
  \bibinfo{volume}{15} (\bibinfo{year}{2016}), \bibinfo{pages}{21:1--21:24}.
\newblock


\bibitem[\protect\citeauthoryear{Jin, Ray, and Edwards}{Jin
  et~al\mbox{.}}{2009}]%
        {jin2009redundant}
\bibfield{author}{\bibinfo{person}{Xin Jin}, \bibinfo{person}{Asok Ray}, {and}
  \bibinfo{person}{Robert~M Edwards}.} \bibinfo{year}{2009}\natexlab{}.
\newblock \showarticletitle{Redundant Sensor Calibration and Estimation for
  Monitoring and Control of Nuclear Power Plants}.
\newblock \bibinfo{journal}{\emph{Transactions of the American Nuclear
  Society}}  \bibinfo{volume}{101} (\bibinfo{year}{2009}),
  \bibinfo{pages}{307--308}.
\newblock


\bibitem[\protect\citeauthoryear{Johansson, V{\"o}lker, Eliasson, {\"O}stmark,
  Lindgren, and Delsing}{Johansson et~al\mbox{.}}{2004}]%
        {johansson2004mulle}
\bibfield{author}{\bibinfo{person}{Jonny Johansson}, \bibinfo{person}{Matthias
  V{\"o}lker}, \bibinfo{person}{Jens Eliasson}, \bibinfo{person}{{\AA}ke
  {\"O}stmark}, \bibinfo{person}{Per Lindgren}, {and} \bibinfo{person}{Jerker
  Delsing}.} \bibinfo{year}{2004}\natexlab{}.
\newblock \showarticletitle{Mulle: a minimal sensor networking device:
  implementation and manufacturing challenges}. In
  \bibinfo{booktitle}{\emph{IMAPS Nordic Annual Conference:
  26/09/2004-28/09/2004}}. International Microelectronics and Packaging
  Society, Nordic chapter, \bibinfo{pages}{265--271}.
\newblock


\bibitem[\protect\citeauthoryear{Kasdorf and Perlman}{Kasdorf and
  Perlman}{2013}]%
        {kasdorf2013hyperthermia}
\bibfield{author}{\bibinfo{person}{Ericalyn Kasdorf} {and}
  \bibinfo{person}{Jeffrey~M Perlman}.} \bibinfo{year}{2013}\natexlab{}.
\newblock \showarticletitle{Hyperthermia, inflammation, and perinatal brain
  injury}.
\newblock \bibinfo{journal}{\emph{Pediatric Neurology}} \bibinfo{volume}{49},
  \bibinfo{number}{1} (\bibinfo{year}{2013}), \bibinfo{pages}{8--14}.
\newblock


\bibitem[\protect\citeauthoryear{Kasmi and Esteves}{Kasmi and Esteves}{2018}]%
        {kasmi2018remote}
\bibfield{author}{\bibinfo{person}{Chaouki Kasmi} {and}
  \bibinfo{person}{Jos{\'e} Esteves}.} \bibinfo{year}{2018}\natexlab{}.
\newblock \showarticletitle{Remote and Silent Voice Command Injection on a
  Smartphone through Conducted IEMI - Threats of Smart IEMI for Information
  Security. Wireless Security Lab, French Network and Information Security
  Agency (ANSSI), Tech. Rep. System Design \& Assessment Note 48}.
\newblock
  \bibinfo{howpublished}{\url{http://ece-research.unm.edu/summa/notes/SDAN/SDAN0048.pdf}}.
\newblock  (\bibinfo{date}{04} \bibinfo{year}{2018}).
\newblock


\bibitem[\protect\citeauthoryear{Kester}{Kester}{1999}]%
        {kester1999practical}
\bibfield{author}{\bibinfo{person}{Walt Kester}.}
  \bibinfo{year}{1999}\natexlab{}.
\newblock \bibinfo{booktitle}{\emph{Practical design techniques for sensor
  signal conditioning}}.
\newblock \bibinfo{publisher}{Analog devices}.
\newblock


\bibitem[\protect\citeauthoryear{Kester, Bryant, and Jung}{Kester
  et~al\mbox{.}}{1999}]%
        {kester1999section}
\bibfield{author}{\bibinfo{person}{Walt Kester}, \bibinfo{person}{James
  Bryant}, {and} \bibinfo{person}{Walt Jung}.} \bibinfo{year}{1999}\natexlab{}.
\newblock \showarticletitle{Temperature sensors (Section 7)}.
\newblock  (\bibinfo{year}{1999}).
\newblock


\bibitem[\protect\citeauthoryear{Khaleghi, Khamis, Karray, and Razavi}{Khaleghi
  et~al\mbox{.}}{2013}]%
        {khaleghi2013multisensor}
\bibfield{author}{\bibinfo{person}{Bahador Khaleghi}, \bibinfo{person}{Alaa
  Khamis}, \bibinfo{person}{Fakhreddine~O Karray}, {and}
  \bibinfo{person}{Saiedeh~N Razavi}.} \bibinfo{year}{2013}\natexlab{}.
\newblock \showarticletitle{Multisensor data fusion: A review of the
  state-of-the-art}.
\newblock \bibinfo{journal}{\emph{Information fusion}} \bibinfo{volume}{14},
  \bibinfo{number}{1} (\bibinfo{year}{2013}), \bibinfo{pages}{28--44}.
\newblock


\bibitem[\protect\citeauthoryear{Kitchin and Counts}{Kitchin and
  Counts}{2004}]%
        {kitchin2004designer}
\bibfield{author}{\bibinfo{person}{Charles Kitchin} {and} \bibinfo{person}{Lew
  Counts}.} \bibinfo{year}{2004}\natexlab{}.
\newblock \bibinfo{booktitle}{\emph{A designer's guide to instrumentation
  amplifiers}}.
\newblock \bibinfo{publisher}{Analog Devices}.
\newblock


\bibitem[\protect\citeauthoryear{Kong, Chen, Xie, and Zhou}{Kong
  et~al\mbox{.}}{2005}]%
        {kong2005distributed}
\bibfield{author}{\bibinfo{person}{Fan-Tian Kong}, \bibinfo{person}{You-Ping
  Chen}, \bibinfo{person}{Jing-Ming Xie}, {and} \bibinfo{person}{Zu-De Zhou}.}
  \bibinfo{year}{2005}\natexlab{}.
\newblock \showarticletitle{Distributed temperature control system based on
  multi-sensor data fusion}. In \bibinfo{booktitle}{\emph{Machine Learning and
  Cybernetics, 2005. Proceedings of 2005 International Conference on}},
  Vol.~\bibinfo{volume}{1}. IEEE, \bibinfo{pages}{494--498}.
\newblock


\bibitem[\protect\citeauthoryear{Kune, Backes, Clark, Kramer, Reynolds, Fu,
  Kim, and Xu}{Kune et~al\mbox{.}}{2013}]%
        {kune2013ghost}
\bibfield{author}{\bibinfo{person}{Denis~Foo Kune}, \bibinfo{person}{John
  Backes}, \bibinfo{person}{Shane~S Clark}, \bibinfo{person}{Daniel Kramer},
  \bibinfo{person}{Matthew Reynolds}, \bibinfo{person}{Kevin Fu},
  \bibinfo{person}{Yongdae Kim}, {and} \bibinfo{person}{Wenyuan Xu}.}
  \bibinfo{year}{2013}\natexlab{}.
\newblock \showarticletitle{Ghost talk: Mitigating EMI signal injection attacks
  against analog sensors}. In \bibinfo{booktitle}{\emph{IEEE Symposium on
  Security and Privacy}}.
\newblock


\bibitem[\protect\citeauthoryear{KV and Smet}{KV and Smet}{2016}]%
        {frameworksensor}
\bibfield{author}{\bibinfo{person}{Santhosh KV} {and} \bibinfo{person}{Karel~De
  Smet}.} \bibinfo{year}{2016}\natexlab{}.
\newblock \showarticletitle{Sensor Data Fusion Framework for Improvement of
  Temperature Sensor Characteristics}.
\newblock \bibinfo{journal}{\emph{Measurement and Control}}
  \bibinfo{volume}{49}, \bibinfo{number}{7} (\bibinfo{year}{2016}),
  \bibinfo{pages}{219--229}.
\newblock


\bibitem[\protect\citeauthoryear{Lee, Gerlach, and Joshi}{Lee
  et~al\mbox{.}}{2008}]%
        {thermalmodeling}
\bibfield{author}{\bibinfo{person}{Jaeho Lee}, \bibinfo{person}{David~W
  Gerlach}, {and} \bibinfo{person}{Yogendra~K Joshi}.}
  \bibinfo{year}{2008}\natexlab{}.
\newblock \showarticletitle{Parametric thermal modeling of heat transfer in
  handheld electronic devices}. In \bibinfo{booktitle}{\emph{Thermal and
  Thermomechanical Phenomena in Electronic Systems, 2008. ITHERM 2008. 11th
  Intersociety Conference on}}. IEEE, \bibinfo{pages}{604--609}.
\newblock


\bibitem[\protect\citeauthoryear{Mance}{Mance}{2008}]%
        {mance2008keeping}
\bibfield{author}{\bibinfo{person}{Martha~J Mance}.}
  \bibinfo{year}{2008}\natexlab{}.
\newblock \showarticletitle{Keeping infants warm: challenges of hypothermia}.
\newblock \bibinfo{journal}{\emph{Advances in neonatal care}}
  \bibinfo{volume}{8}, \bibinfo{number}{1} (\bibinfo{year}{2008}),
  \bibinfo{pages}{6--12}.
\newblock


\bibitem[\protect\citeauthoryear{Mary and Eamonn}{Mary and Eamonn}{2006}]%
        {mccarthyadc}
\bibfield{author}{\bibinfo{person}{McCarthy Mary} {and} \bibinfo{person}{Dillon
  Eamonn}.} \bibinfo{year}{2006}\natexlab{}.
\newblock \bibinfo{title}{{ADC Requirements for Temperature Measurement
  Systems}}.
\newblock
  \bibinfo{howpublished}{\url{https://www.analog.com/media/en/technical-documentation/application-notes/AN-880.pdf?doc=UG-181.pdf}}.
\newblock


\bibitem[\protect\citeauthoryear{{Mishra}, {Potnis}, {Dwivedy}, and
  {Meena}}{{Mishra} et~al\mbox{.}}{2017}]%
        {sdr}
\bibfield{author}{\bibinfo{person}{M. {Mishra}}, \bibinfo{person}{A. {Potnis}},
  \bibinfo{person}{P. {Dwivedy}}, {and} \bibinfo{person}{S.~K. {Meena}}.}
  \bibinfo{year}{2017}\natexlab{}.
\newblock \showarticletitle{Software defined radio based receivers using
  RTL-SDR: A review}. In \bibinfo{booktitle}{\emph{2017 International
  Conference on Recent Innovations in Signal processing and Embedded Systems
  (RISE)}}.
\newblock


\bibitem[\protect\citeauthoryear{Molgat-Seon, Daboval, Chou, and
  Jay}{Molgat-Seon et~al\mbox{.}}{2013}]%
        {molgat2013accidental}
\bibfield{author}{\bibinfo{person}{Y Molgat-Seon}, \bibinfo{person}{T Daboval},
  \bibinfo{person}{S Chou}, {and} \bibinfo{person}{O Jay}.}
  \bibinfo{year}{2013}\natexlab{}.
\newblock \showarticletitle{Accidental overheating of a newborn under an infant
  radiant warmer: a lesson for future use}.
\newblock \bibinfo{journal}{\emph{Journal of Perinatology}}
  \bibinfo{volume}{33}, \bibinfo{number}{9} (\bibinfo{year}{2013}),
  \bibinfo{pages}{738}.
\newblock


\bibitem[\protect\citeauthoryear{Morrison}{Morrison}{1977}]%
        {morrison}
\bibfield{author}{\bibinfo{person}{Ralph Morrison}.}
  \bibinfo{year}{1977}\natexlab{}.
\newblock \bibinfo{booktitle}{\emph{Grounding and shielding techniques in
  instrumentation}}.
\newblock \bibinfo{publisher}{Wiley New York}.
\newblock


\bibitem[\protect\citeauthoryear{Park, Son, Shin, Kim, and Kim}{Park
  et~al\mbox{.}}{2016}]%
        {park2016ain}
\bibfield{author}{\bibinfo{person}{Youngseok Park}, \bibinfo{person}{Yunmok
  Son}, \bibinfo{person}{Hocheol Shin}, \bibinfo{person}{Dohyun Kim}, {and}
  \bibinfo{person}{Yongdae Kim}.} \bibinfo{year}{2016}\natexlab{}.
\newblock \showarticletitle{This Ain{\textquoteright}t Your Dose: Sensor
  Spoofing Attack on Medical Infusion Pump}. In \bibinfo{booktitle}{\emph{10th
  {USENIX} Workshop on Offensive Technologies ({WOOT})}}.
\newblock


\bibitem[\protect\citeauthoryear{Payne}{Payne}{2016}]%
        {survival_rate1}
\bibfield{author}{\bibinfo{person}{Elizabeth Payne}.}
  \bibinfo{year}{2016}\natexlab{}.
\newblock \bibinfo{title}{A Brief History of Advances in Neonatal Care}.
\newblock
  \bibinfo{howpublished}{\url{https://www.nicuawareness.org/blog/a-brief-history-of-advances-in-neonatal-care}}.
\newblock


\bibitem[\protect\citeauthoryear{Petit, Stottelaar, and Feiri}{Petit
  et~al\mbox{.}}{2015}]%
        {Petit2015RemoteAO}
\bibfield{author}{\bibinfo{person}{Jonathan Petit}, \bibinfo{person}{Bas
  Stottelaar}, {and} \bibinfo{person}{Michael Feiri}.}
  \bibinfo{year}{2015}\natexlab{}.
\newblock \showarticletitle{Remote Attacks on Automated Vehicles Sensors :
  Experiments on Camera and LiDAR}.
\newblock


\bibitem[\protect\citeauthoryear{Poulton}{Poulton}{1994}]%
        {emiopamp}
\bibfield{author}{\bibinfo{person}{AS Poulton}.}
  \bibinfo{year}{1994}\natexlab{}.
\newblock \showarticletitle{{Effect of conducted EMI on the DC performance of
  operational amplifiers}}.
\newblock \bibinfo{journal}{\emph{Electronics letters}} \bibinfo{volume}{30},
  \bibinfo{number}{4} (\bibinfo{year}{1994}), \bibinfo{pages}{282--284}.
\newblock


\bibitem[\protect\citeauthoryear{Ray and Luck}{Ray and Luck}{1991}]%
        {ray1991introduction}
\bibfield{author}{\bibinfo{person}{Asok Ray} {and} \bibinfo{person}{Rogelio
  Luck}.} \bibinfo{year}{1991}\natexlab{}.
\newblock \showarticletitle{An introduction to sensor signal validation in
  redundant measurement systems}.
\newblock \bibinfo{journal}{\emph{IEEE Control Systems}} \bibinfo{volume}{11},
  \bibinfo{number}{2} (\bibinfo{year}{1991}), \bibinfo{pages}{44--49}.
\newblock


\bibitem[\protect\citeauthoryear{Ross-Pinnock and Maropoulos}{Ross-Pinnock and
  Maropoulos}{2016}]%
        {thermocouple}
\bibfield{author}{\bibinfo{person}{David Ross-Pinnock} {and}
  \bibinfo{person}{Paul~G Maropoulos}.} \bibinfo{year}{2016}\natexlab{}.
\newblock \showarticletitle{Review of industrial temperature measurement
  technologies and research priorities for the thermal characterization of the
  factories of the future}. In \bibinfo{booktitle}{\emph{Proceedings of the
  Institution of Mechanical Engineers, Part B: Journal of Engineering
  Manufacture}}, Vol.~\bibinfo{volume}{230}. \bibinfo{pages}{793--806}.
\newblock


\bibitem[\protect\citeauthoryear{Selvaraj}{Selvaraj}{2018}]%
        {selvaraj2018intentional}
\bibfield{author}{\bibinfo{person}{Jayaprakash Selvaraj}.}
  \bibinfo{year}{2018}\natexlab{}.
\newblock \emph{\bibinfo{title}{Intentional Electromagnetic Interference Attack
  on Sensors and Actuators}}.
\newblock \bibinfo{thesistype}{Ph.D. Dissertation}. \bibinfo{school}{Iowa State
  University}.
\newblock


\bibitem[\protect\citeauthoryear{Selvaraj, Dayan{\i}kl{\i}, Gaunkar, Ware,
  Gerdes, Mina, et~al\mbox{.}}{Selvaraj et~al\mbox{.}}{2018}]%
        {selvaraj2018electromagnetic}
\bibfield{author}{\bibinfo{person}{Jayaprakash Selvaraj},
  \bibinfo{person}{G{\"o}k{\c{c}}en~Y Dayan{\i}kl{\i}},
  \bibinfo{person}{Neelam~Prabhu Gaunkar}, \bibinfo{person}{David Ware},
  \bibinfo{person}{Ryan~M Gerdes}, \bibinfo{person}{Mani Mina},
  {et~al\mbox{.}}} \bibinfo{year}{2018}\natexlab{}.
\newblock \showarticletitle{Electromagnetic Induction Attacks Against Embedded
  Systems}. In \bibinfo{booktitle}{\emph{Proceedings of the 2018 on Asia
  Conference on Computer and Communications Security}}.
  \bibinfo{pages}{499--510}.
\newblock


\bibitem[\protect\citeauthoryear{Shin, Kim, Kwon, and Kim}{Shin
  et~al\mbox{.}}{2017}]%
        {shin2017illusion}
\bibfield{author}{\bibinfo{person}{Hocheol Shin}, \bibinfo{person}{Dohyun Kim},
  \bibinfo{person}{Yujin Kwon}, {and} \bibinfo{person}{Yongdae Kim}.}
  \bibinfo{year}{2017}\natexlab{}.
\newblock \showarticletitle{Illusion and Dazzle: Adversarial Optical Channel
  Exploits against Lidars for Automotive Applications}. In
  \bibinfo{booktitle}{\emph{International Conference on Cryptographic Hardware
  and Embedded Systems}}. Springer.
\newblock


\bibitem[\protect\citeauthoryear{Son, Shin, Kim, Park, Noh, Choi, Choi, and
  Kim}{Son et~al\mbox{.}}{2015}]%
        {son2015rocking}
\bibfield{author}{\bibinfo{person}{Yunmok Son}, \bibinfo{person}{Hocheol Shin},
  \bibinfo{person}{Dongkwan Kim}, \bibinfo{person}{Youngseok Park},
  \bibinfo{person}{Juhwan Noh}, \bibinfo{person}{Kibum Choi},
  \bibinfo{person}{Jungwoo Choi}, {and} \bibinfo{person}{Yongdae Kim}.}
  \bibinfo{year}{2015}\natexlab{}.
\newblock \showarticletitle{Rocking Drones with Intentional Sound Noise on
  Gyroscopic Sensors}. In \bibinfo{booktitle}{\emph{Proceedings of USENIX
  Security Symposium}}.
\newblock


\bibitem[\protect\citeauthoryear{{Stagner}, {Conrad}, {Osterwise}, {Beetner},
  and {Grant}}{{Stagner} et~al\mbox{.}}{2011}]%
        {super}
\bibfield{author}{\bibinfo{person}{C. {Stagner}}, \bibinfo{person}{A.
  {Conrad}}, \bibinfo{person}{C. {Osterwise}}, \bibinfo{person}{D.~G.
  {Beetner}}, {and} \bibinfo{person}{S. {Grant}}.}
  \bibinfo{year}{2011}\natexlab{}.
\newblock \showarticletitle{A Practical Superheterodyne-Receiver Detector Using
  Stimulated Emissions}.
\newblock \bibinfo{journal}{\emph{IEEE Transactions on Instrumentation and
  Measurement}} \bibinfo{volume}{60}, \bibinfo{number}{4}
  (\bibinfo{year}{2011}), \bibinfo{pages}{1461--1468}.
\newblock


\bibitem[\protect\citeauthoryear{Tilson, Field, et~al\mbox{.}}{Tilson
  et~al\mbox{.}}{2006}]%
        {overheat_death}
\bibfield{author}{\bibinfo{person}{Hugh Tilson}, \bibinfo{person}{Marilyn~J
  Field}, {et~al\mbox{.}}} \bibinfo{year}{2006}\natexlab{}.
\newblock \bibinfo{booktitle}{\emph{Safe Medical Devices for Children. Chapter:
  4 Identifying and Understanding Adverse Medical Device Events. (Page
  147-148)}}.
\newblock \bibinfo{publisher}{National Academies Press}.
\newblock


\bibitem[\protect\citeauthoryear{Trippel, Weisse, Xu, Honeyman, and Fu}{Trippel
  et~al\mbox{.}}{2017}]%
        {trippel2017walnut}
\bibfield{author}{\bibinfo{person}{Timothy Trippel}, \bibinfo{person}{Ofir
  Weisse}, \bibinfo{person}{Wenyuan Xu}, \bibinfo{person}{Peter Honeyman},
  {and} \bibinfo{person}{Kevin Fu}.} \bibinfo{year}{2017}\natexlab{}.
\newblock \showarticletitle{WALNUT: Waging doubt on the integrity of {MEMS}
  accelerometers with acoustic injection attacks}. In
  \bibinfo{booktitle}{\emph{Proceedings of IEEE European Symposium on Security
  and Privacy}}.
\newblock


\bibitem[\protect\citeauthoryear{Tu, Lin, Lee, and Hei}{Tu
  et~al\mbox{.}}{2018}]%
        {tu18injected}
\bibfield{author}{\bibinfo{person}{Yazhou Tu}, \bibinfo{person}{Zhiqiang Lin},
  \bibinfo{person}{Insup Lee}, {and} \bibinfo{person}{Xiali Hei}.}
  \bibinfo{year}{2018}\natexlab{}.
\newblock \showarticletitle{Injected and Delivered: {Fabricating} Implicit
  Control over Actuation Systems by Spoofing Inertial Sensors}. In
  \bibinfo{booktitle}{\emph{Proceedings of USENIX Security Symposium}}.
\newblock


\bibitem[\protect\citeauthoryear{Walter and Carraretto}{Walter and
  Carraretto}{2016}]%
        {walter2016neurological}
\bibfield{author}{\bibinfo{person}{Edward~James Walter} {and}
  \bibinfo{person}{Mike Carraretto}.} \bibinfo{year}{2016}\natexlab{}.
\newblock \showarticletitle{The neurological and cognitive consequences of
  hyperthermia}.
\newblock \bibinfo{journal}{\emph{Critical Care}} \bibinfo{volume}{20},
  \bibinfo{number}{1} (\bibinfo{year}{2016}), \bibinfo{pages}{199}.
\newblock


\bibitem[\protect\citeauthoryear{Wang and Lee}{Wang and Lee}{2010}]%
        {wang2010analysis}
\bibfield{author}{\bibinfo{person}{Shuo Wang} {and} \bibinfo{person}{Fred~C
  Lee}.} \bibinfo{year}{2010}\natexlab{}.
\newblock \showarticletitle{Analysis and applications of parasitic capacitance
  cancellation techniques for EMI suppression}.
\newblock \bibinfo{journal}{\emph{IEEE Transactions on Industrial Electronics}}
  \bibinfo{volume}{57}, \bibinfo{number}{9} (\bibinfo{year}{2010}),
  \bibinfo{pages}{3109--3117}.
\newblock


\bibitem[\protect\citeauthoryear{Wang, Wang, Yang, Li, and Pan}{Wang
  et~al\mbox{.}}{2017}]%
        {wang2017sonic}
\bibfield{author}{\bibinfo{person}{Zhengbo Wang}, \bibinfo{person}{Kang Wang},
  \bibinfo{person}{Bo Yang}, \bibinfo{person}{Shangyuan Li}, {and}
  \bibinfo{person}{Aimin Pan}.} \bibinfo{year}{2017}\natexlab{}.
\newblock \showarticletitle{Sonic gun to smart devices: Your devices lose
  control under ultrasound/sound}.
\newblock \bibinfo{journal}{\emph{BlackHat USA}} (\bibinfo{year}{2017}).
\newblock


\bibitem[\protect\citeauthoryear{Weber, Schinkel, Hoene, Guttowski, John, and
  Reichl}{Weber et~al\mbox{.}}{2005}]%
        {weber2005radio}
\bibfield{author}{\bibinfo{person}{S Weber}, \bibinfo{person}{M Schinkel},
  \bibinfo{person}{E Hoene}, \bibinfo{person}{S Guttowski}, \bibinfo{person}{W
  John}, {and} \bibinfo{person}{H Reichl}.} \bibinfo{year}{2005}\natexlab{}.
\newblock \showarticletitle{Radio frequency characteristics of high power
  common-mode chokes}. In \bibinfo{booktitle}{\emph{IEEE Int. Zurich Symp. on
  Electromagnetic Compatibility}}. \bibinfo{pages}{1--4}.
\newblock


\bibitem[\protect\citeauthoryear{Welte}{Welte}{2007}]%
        {vaccine_spoil}
\bibfield{author}{\bibinfo{person}{Melanie Welte}.}
  \bibinfo{year}{2007}\natexlab{}.
\newblock \bibinfo{title}{USA Today: Vaccines ruined by poor refrigeration}.
\newblock
  \bibinfo{howpublished}{\url{https://usatoday30.usatoday.com/news/health/2007-12-04-spoiled-vaccines_N.htm}}.
\newblock


\bibitem[\protect\citeauthoryear{Weston}{Weston}{2016}]%
        {weston2016electromagnetic}
\bibfield{author}{\bibinfo{person}{David Weston}.}
  \bibinfo{year}{2016}\natexlab{}.
\newblock \bibinfo{booktitle}{\emph{Electromagnetic Compatibility: Methods,
  Analysis, Circuits, and Measurement}}.
\newblock \bibinfo{publisher}{Crc Press}.
\newblock


\bibitem[\protect\citeauthoryear{Wu, Li, Pommerenke, Khilkevich, and Hess}{Wu
  et~al\mbox{.}}{2018}]%
        {wu2018characterization}
\bibfield{author}{\bibinfo{person}{Chunyu Wu}, \bibinfo{person}{Guanghua Li},
  \bibinfo{person}{David~J Pommerenke}, \bibinfo{person}{Victor Khilkevich},
  {and} \bibinfo{person}{Gary Hess}.} \bibinfo{year}{2018}\natexlab{}.
\newblock \showarticletitle{Characterization of the RFI Rectification Behavior
  of Instrumentation Amplifiers}. In \bibinfo{booktitle}{\emph{2018 IEEE
  Symposium on Electromagnetic Compatibility, Signal Integrity and Power
  Integrity (EMC, SI \& PI)}}. IEEE, \bibinfo{pages}{156--160}.
\newblock


\bibitem[\protect\citeauthoryear{Yan, Xu, and Liu}{Yan et~al\mbox{.}}{2016}]%
        {yan2016can}
\bibfield{author}{\bibinfo{person}{Chen Yan}, \bibinfo{person}{Wenyuan Xu},
  {and} \bibinfo{person}{Jianhao Liu}.} \bibinfo{year}{2016}\natexlab{}.
\newblock \showarticletitle{Can you trust autonomous vehicles: Contactless
  attacks against sensors of self-driving vehicle}.
\newblock \bibinfo{journal}{\emph{DEF CON}}  \bibinfo{volume}{24}
  (\bibinfo{year}{2016}).
\newblock


\bibitem[\protect\citeauthoryear{Zhang, Yan, Ji, Zhang, Zhang, and Xu}{Zhang
  et~al\mbox{.}}{2017}]%
        {zhang2017dolphinattack}
\bibfield{author}{\bibinfo{person}{Guoming Zhang}, \bibinfo{person}{Chen Yan},
  \bibinfo{person}{Xiaoyu Ji}, \bibinfo{person}{Tianchen Zhang},
  \bibinfo{person}{Taimin Zhang}, {and} \bibinfo{person}{Wenyuan Xu}.}
  \bibinfo{year}{2017}\natexlab{}.
\newblock \showarticletitle{DolphinAttack: Inaudible voice commands}. In
  \bibinfo{booktitle}{\emph{Proceedings of the 2017 ACM SIGSAC Conference on
  Computer and Communications Security}}.
\newblock


\end{thebibliography}

\appendix  \vspace{-2mm}
\section*{Appendix} 

\subsection*{The Setup of DPI Experiments on the Experimental Temperature Sensing Circuitry} \vspace{-2mm}

\begin{figure}[h]
\centering
 \includegraphics[scale=0.58]{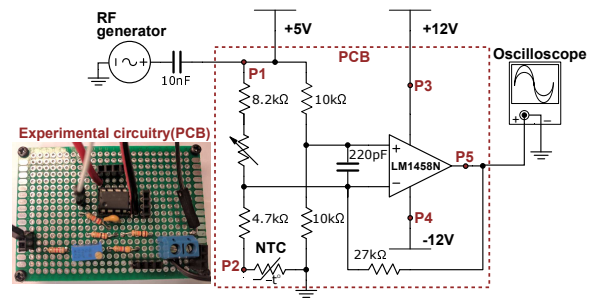}
  \vspace{-3mm}
  \caption{The setup of direct power injections through different injection points of a typical NTC-based temperature sensing circuitry. In this illustration, the signal injection circuit is connected to the injection point P1.}\label{fig_ntc_circuits}
\vspace{-4mm}
\end{figure}

As shown in Fig. \ref{fig_ntc_circuits}, a 1-meter NTC thermistor is wired in a bridge circuit, which is excited by a $+5V$ DC power source.
The differential voltage generated by the bridge circuit is collected and amplified by a Texas Instruments (TI) LM1458 operational amplifier. By measuring the output voltage of the amplifier, the resistance of the thermistor and the corresponding temperature can be calculated. We choose the circuit elements based on the schematics of temperature sensing circuits in infant incubators \cite{incubator1,incubator2}. During the experiments, the $+5V$ and $\pm 12V$ DC voltage sources are provided by an Agilent E3630A triple output power supply. We monitor the analog amplified output with an Agilent MSOX4054A oscilloscope.

In our DPI experiments, we connect the output of the signal injection circuit to each of the signal injection points on the sensing circuitry. A $10nF$ capacitor is used to decouple the DC signal in the experimental circuitry from the signal injection circuitry. The source of the EMI signals is an Agilent N5172B vector signal generator.

\end{document}